\documentclass[useAMS,usenatbib]{emulateapj}

\usepackage{comment}
\usepackage{amsmath}
\newcommand\msun {M$_{\odot}$}
\def\approxgt{\ifmmode \rlap{$>$}{}_{{}_{{}_{\textstyle\sim}}} \else%
$\rlap{$>$}{}_{{}_{{}_{\textstyle\sim}}}$\fi} 
\def\approxlt{\ifmmode \rlap{$<$}{}_{{}_{{}_{\textstyle\sim}}} \else%
$\rlap{$<$}{}_{{}_{{}_{\textstyle\sim}}}$\fi}

\def\arcsec{\hbox{$^{\prime\prime}$}}

\def\flx{erg cm$^{-2}$ s$^{-1}$}
\def\lum{erg s$^{-1}$}
\def\xmm{{\it XMM--Newton}}

\def\srcone{PTF09axc}
\def\srctwo{PTF09ge}
\def\srcthree{ASASSN--14ae}
\def\srcfour{PTF09djl}
\def\chan{{\it Chandra}}

\shorttitle{Late--Time X--ray Detections of Optically Selected TDEs}
\shortauthors{Jonker et al.}

\begin{document}
\title{Implications from Late--Time X--ray Detections of Optically Selected Tidal Disruption Events: State Changes, Unification, and Detection Rates}

\author{P.G.~Jonker \altaffilmark{1,2}} 
\email{p.jonker@sron.nl}

\author{N.C.~Stone\altaffilmark{3,4,5}} 
\author{A.~Generozov\altaffilmark{5,6}}
\author{S.~van Velzen\altaffilmark{7,8}}
\author{B.~Metzger\altaffilmark{5}}

\altaffiltext{1}{SRON, Netherlands Institute for Space Research, Sorbonnelaan 2,
  3584~CA, Utrecht, The Netherlands}
\altaffiltext{2}{Department of Astrophysics/IMAPP, Radboud University,
P.O.~Box 9010, 6500 GL, Nijmegen, The Netherlands}
\altaffiltext{3}{Department of Astronomy, University of Maryland, Stadium Drive, College Park, MD 20742-2421, U.S.A.}
\altaffiltext{4}{Racah Institute of Physics, The Hebrew University, Jerusalem, 91904, Israel}
\altaffiltext{5}{Department of Physics, Department of Astronomy, and Columbia Astrophysics Laboratory, Columbia University, New York, NY 10027, USA}
\altaffiltext{6}{JILA, University of Colorado and National Institute of Standards and Technology, 440 UCB, Boulder, CO 80309-0440, USA}
\altaffiltext{7}{Center for Cosmology and Particle Physics, New York University, New York, NY 10003, U.S.A.}
\altaffiltext{8}{Joint Space-Science Institute, University of Maryland, College Park, MD 20742, U.S.A.}

\begin{abstract}
We present Chandra X-ray observations of four optically-selected tidal
disruption events (TDEs) obtained 4--9 years after discovery. Three sources were
detected with luminosities between $9\times 10^{40}$ and $3\times 10^{42}$\lum.
The spectrum of \srcone\, is consistent with a power law of index 2.5$\pm$0.1,
whereas the spectrum of \srctwo\, is very soft. The power law spectrum of
\srcone\, and prior literature findings, provide evidence that TDEs transition
from an early-time soft state to a late-time hard state many years after
disruption. We propose that the time to peak luminosity for optical and X-ray
emission may differ substantially in TDEs, with X-rays being produced or
becoming observable later. This delay helps explain the differences in observed
properties such as L$_{\rm opt}/$L$_{\rm X}$ of optically and X-ray selected
TDEs. We update TDE rate predictions for the eROSITA instrument: it ranges from
$3~{\rm yr}^{-1}$ to $990~{\rm yr}^{-1}$, depending sensitively on the
distribution of black hole spins and the time delay between disruption and peak
X-ray brightness. We further predict an asymmetry in the number of retrograde
and prograde disks in samples of optically and X--ray selected TDEs. The details
of the observational biases can contribute to observed differences between
optically and X-ray selected TDEs (with optically selected TDEs being fainter in
X-rays for retrograde TDE disks).

\end{abstract}


\section{Introduction}

Stellar tidal disruption is an unavoidable outcome of collisional orbital dynamics in dense stellar systems \citep{1976MNRAS.176..633F}.  The stochastic two--body relaxation of orbital parameters leads stars on a random walk through angular momentum space, eventually delivering them to pericenters close to the supermassive black hole (SMBH). Once a star's orbital pericenter falls within the tidal, or Roche, radius of the SMBH, the star will be destroyed upon pericenter passage
\citep{1975Natur.254..295H, Rees88}.  The resulting tidal disruption events (TDEs) were theoretical curiosities for many years, but have been discovered in increasing numbers over the last two decades.  There are now dozens of known TDEs discovered as transient nuclear flares, which have been identified primarily through quasi--thermal emission in soft X-ray \citep[e.g.,][]{1996A&A...309L..35B, Greiner+00,  2004ApJ...603L..17K,2014ApJ...781...59D}, UV \citep{2006ApJ...653L..25G,2009ApJ...698.1367G}, and optical \citep{2011ApJ...741...73V,2012Natur.485..217G,Chornock+14, 2014ApJ...793...38A, 2014MNRAS.445.3263H,2016MNRAS.455.2918H,2016MNRAS.463.3813H, vvelzen2019-ztf} wavelengths.  A minority of TDEs have been observed to launch relativistic jets detectable (via non--thermal hard X--ray and soft $\gamma$--ray emission) to cosmological distances (e.g.~\citealt{2011Sci...333..203B}; \citealt{2011Sci...333..199L}).  However, late--time radio followup of thermally--selected TDEs usually returns upper limits \citep{2013A&A...552A...5V, Bower+13}, suggesting that only a minority of TDEs are accompanied by very high luminosity jets \citep{Generozov+17}.

Astrophysical interest in TDEs is manifold.  These flares hold great scientific potential as probes of SMBH demographics, as the mass fallback rate onto the black hole encodes the mass \citep{Rees88, Lodato+09, Guillochon&RamirezRuiz13} of the SMBH.  The SMBH spin may be more subtly imprinted into TDE observables \citep{Stone&Loeb12, Guillochon&RamirezRuiz15, Hayasaki+16}.  In the subset of TDEs that launch relativistic jets, radio synchrotron emission produced in the jet forward shock can place tight constraints on circumnuclear gas in distant galactic nuclei \citep{Giannios&Metzger11, 2012ApJ...748...36B}.  More speculatively, these jets could be responsible for the observed flux of ultra--high energy cosmic rays \citep{2009ApJ...693..329F, Farrar&Piran14}.  Exotic TDEs may serve as signposts of unusual SMBH dynamics: truncated
light curves are expected in the vicinity of close SMBH binaries
\citep{2009ApJ...706L.133L}, and off-nuclear TDEs may indicate SMBHs recoiling
after anisotropic gravitational wave emission (\citealt{2011MNRAS.412...75S}; \citealt{Jonker2012}). Finally, TDEs may also serve as natural accretion physics laboratories, as the mass fallback feeding the disk declines from super--Eddington levels to a few percent of Eddington over a period of months to years \citep{Shen&Matzner14}.  As TDE accretion rates decline from super--Eddington, to modestly
sub--Eddington, to very sub--Eddington levels, their accretion disks might exhibit state changes analogous to those
of stellar--mass black holes in X--ray binaries (XRBs; \citealt{2004MNRAS.355.1105F}; \citealt{2004ApJ...603L..17K}).

Early models for TDE light curves and spectra assumed that the highly eccentric debris streams from stellar disruption would quickly circularize into a compact accretion disk \citep{Rees88, Cannizzo+90, Ulmer99} that might resemble a scaled--up XRB disk, or the innermost regions of an active galactic nucleus (AGN).  A circularized TDE disk would differ from both of these analogues in its radial extent: typically, the tidal radius $R_{\rm t} \lesssim 100 R_{\rm g}$, where $R_{\rm g}$ is the SMBH gravitational radius; a scale much smaller than the typical XRB or AGN disk.

This simple expectation has, however, been strongly challenged.  Recent analytic and numerical theory has found that circularization may be very slow if the debris pericenter $R_{\rm p} \gg 10 R_{\rm g}$ \citep{Shiokawa+15, Dai+15, Piran+15} and/or there is strong misalignment between the SMBH spin vector and the debris angular momentum vector \citep{Guillochon&RamirezRuiz15, Hayasaki+16}.  In tandem, early--time observations have found four properties characteristic of optical/UV-selected TDEs (\citealt{2011ApJ...741...73V}; \citealt{2014ApJ...793...38A}; \citealt{2018ApJS..238...15H}): \newline {\it (i)}
low blackbody temperatures ($T_{\rm BB} \approx 2 \times 10^4~{\rm K}$) with blackbody radii
$R_{\rm BB} \sim 10^{2-3} R_{\rm g}$, {\it (ii)} little cooling
(${\rm d} \ln(T_{\rm BB})/{\rm d}t<0.01$~day$^{-1}$) over a $\sim$100 day baseline, {\it
  (iii)} a steep power--law decay in observed flux $F(t)$ often consistent with
$F\propto t^{-5/3}$, and {\it (iv)} very high optical/UV luminosities, with $L_{\rm BB} \sim 10^{43.5-44.5}~{\rm erg~s}^{-1}$ near peak.  

All of these properties are inconsistent with the simplest TDE emission model, which assumes emission from radii $\lesssim R_{\rm t} \sim 10 R_{\rm g}$ \citep{Ulmer99}.  In this scenario, the optical/UV emission is far down the Raleigh--Jeans tail
of the disk spectral energy distribution (SED), and therefore decays slowly in time, $L_{\rm RJ} \propto T_{\rm BB} \propto t^{-5/12}$ \citep{Lodato&Rossi11}.  The predicted level of optical/UV luminosity is $L_{\rm opt} \sim 10^{41}~{\rm erg~s}^{-1}$, far lower than observed.  These discrepancies have motivated multiple theoretical alternatives for the observed optical/UV emission: photon--driven \citep{2009MNRAS.400.2070S} or line--driven
\citep{Miller15} outflows; emission powered by shocks at debris
stream self-intersections \citep{Piran+15}; or thermal reprocessing of accretion power
by a layer of gas at large radii \citep{Loeb&Ulmer97, 2014ApJ...783...23G}. 

Conversely, soft X--ray observations of TDEs are more qualitatively consistent with the simple picture of a compact accretion disk.  Most X--ray detections of TDEs find very soft spectra, consistent with the Wien tail of (multi--color) black bodies at temperatures $T \lesssim 0.1\,{\rm keV}$
\citep{Auchettl+17}, like a scaled--up version of a high-soft
state XRB.  However, these X-ray spectra are almost always taken in the first one or two years of the
flare, when accretion rates are expected to be, at the very least, at a large fraction of the Eddington limit.  Notably, many optically selected TDEs go undetected in X--rays \citep{2012Natur.485..217G}
and, vice versa, X--ray selected TDEs often lack optical variability. For instance, the TDE XMMSL1~J074008.2$-$853927 reported by \citet{2017A&A...598A..29S} does not show a large enhancement in the optical. Some even show no evidence for enhanced optical emission. For instance, the TDE SDSS~J120136.02$+$300305.5 discovered by \citet{2012A&A...541A.106S} had an X--ray luminosity of 3$\times 10^{44}$\lum\, at discovery while the optical spectrum obtained 12 days after the X--ray discovery shows no spectroscopic features (such as broad emission lines) that are usually associated with TDEs. A recent X--ray discovered source, XMMSL2~J144605.0$+$685735 (\citealt{2019A&A...630A..98S}), also shows little or no optical emission above the contribution of the nuclear region of the host galaxy. 

So far, we have discussed the state of the art in {\it early--time} TDE observations, by which we mean observations taken within two years of the peak of the flare.  The behavior of TDE disks at late times is relatively under--explored. We note two differences between the early-- and late--time phases:
\begin{enumerate}
    \item The large theoretical uncertainties associated with circularization and disk formation will be less important long after the peak of the mass return rate.  A quasi--circular disk is a more reasonable approximation at late times, even if initial circularization was inefficient due to weak apsidal precession \citep{Shiokawa+15} or misaligned SMBH spin \citep{Guillochon&RamirezRuiz15, Hayasaki+16}.
    \item The monotonically declining debris fallback rate suggests that at sufficiently late times, TDE disks may pass through the range of sub-Eddington accretion rates that produces a state change in XRB disks (e.g.~\citealt{2011MNRAS.417L..51V}; \citealt{Giannios&Metzger11}; \citealt{Tchekhovskoy+14}).  This analogy suggests that once TDE accretion rates decline below a few percent of Eddington, X--ray emission may exhibit features of the XRB low/hard state, such as a primarily non--thermal, hard power--law spectrum.  Such ``SMBH state changes'' have not yet been seen in TDEs, although there is one suggestive example: X-ray observations of the TDE in IC~3599 show a transition from a soft to harder spectrum at late times (\citealt{1999A&A...343..775K}).
\end{enumerate}
The search for late--time TDE X--ray emission is further motivated by the recent {\it Hubble
Space Telescope} discovery of late--time FUV emission in six optically--selected TDEs \citep{vanVelzen+18}.  In all six cases, the late--time FUV luminosities were well above the levels predicted from extrapolating a naive $\propto t^{-5/3}$ power--law.  The observed slower rate of decline hints at a transition from fallback--dominated to disk--dominated accretion rates \citep{Cannizzo+90}, and the small fitted black body radii ($R_{\rm BB} \sim 2-5 R_{\rm t}$) 
indicate that if optically thick reprocessing layers once existed, they have since dissipated.  It is therefore reasonable to expect that many optically--selected TDEs should, at late times, be emitting relatively unobscured X--rays from their inner disks.

In this paper, we present and analyze \chan\, observations of four optically--selected TDEs taken at late times, long after the peak of the optical flare has passed.  We have observed 
\srcone{} and \srctwo{} 8 years after their discovery, \srcfour{} 9 years after its discovery, and \srcthree{} 5 years after its discovery.  In \S \ref{sec:obs}, we present our observations and results, and in \S \ref{sec:disc}, we discuss the implications of both detections and non--detections for broader questions in TDE and accretion physics.  We adopt $\Omega_m= 0.3$, $\Omega_\Lambda=0.7$, and
$H_0 = 70$ km s$^{-1}$ Mpc$^{-1}$ to convert the redshift of each source to
a luminosity distance.

\section{Observations, analysis and results}
\label{sec:obs}

We obtained 69.19, 34.15, 9.6, 19.08 ksec long on--source \chan\,
exposures of \srcone, \srctwo, \srcthree, and \srcfour,
respectively. The first two sources were observed under {\it Chandra} Guest Observer program 18700591, and the latter two under 20700515.  The observation of \srcone\, was split in two parts of
53.66 and 15.53 ksec in length. The observation identification (ID)
numbers for the data presented here are 19532 (53.66 ksec) and 20879
(15.53 ksec) for \srcone, 19531 for \srctwo, 21503 for \srcthree, and
21504 for \srcfour\, with observing dates and start times (UTC) of
2017-12-08 at 23:11:32, 2017-12-06 at 18:12:18, 2017-09-28 at
20:19:15, 2018-11-17 at 21:48:37, and 2019-01-06 at 13:08:18,
respectively. A log of the observations can be found in 
Table~\ref{tab:log}.

\begin{table*}
  \caption{A log of the \chan\, late--time X--ray observations of four 
  optically selected tidal disruption events. The time since the discovery of the optical transient is denoted with $\Delta t$ (delay).}
\label{tab:log}
\begin{center}
\begin{tabular}{ccccc}
\hline
Source & Observing date & Observation ID & Duration & Delay \\
       &      MJD (UTC) &           & (kilo seconds) & ($\Delta t$; yr) \\
\hline
\srcone & 58095.966 &  19532 & 53.66 & 8.5 \\
\srcone & 58093.759 & 20879 & 15.53 & 8.5 \\
\srctwo & 58024.847 & 19531 & 34.15 & 8.4 \\
\srcthree & 58439.909 & 21503 & 9.6 & 4.8 \\
\srcfour & 58489.547 & 21504 & 19.08 & 9.5 \\
\end{tabular}
\end{center}
\end{table*}

In all cases, the source position as derived in the initial optical
outburst was covered by the S3 CCD of the ACIS-S detector array
(\citealt{1997AAS...190.3404G}). For the observations of \srcone\, and
\srctwo, 3 CCDs were operational (besides the S3 CCD, S4 and S2 were
operational) and the full CCDs were read out providing a nominal
exposure time per frame of 3.1~sec. For the observations of
\srcthree\, and \srcfour\, we chose to use only the S3 CCD. It was
operated in sub--array mode where only a quarter of the CCD is read out. This yields an
exposure time of 0.8~s per CCD frame.

We reprocessed and analyzed the data using the {\sc ciao} 4.10
software developed by the \chan\, X--ray Center and employing {\sc
  caldb} version 4.8.1. To allow for a thorough rejection of events
unrelated to the source such as cosmic ray hits, the data telemetry
mode was set to {\it very faint}. Using the {\sc ciao} tool {\it
  wavdetect} we have detected an X--ray source in an image constructed
from the 0.3--7 keV data. The position of the X-ray source is
consistent with the optical position of the TDE in all three cases
where we detected a source close to the expected position 
(see Table \ref{tab:coor}). No X--ray source was detected at the 
location of the optical outburst source in the case of \srcfour.

For the detected sources we calculate the 95\% confidence 
uncertainty on the \chan\, X--ray position using eq.~12 in \citet{2007ApJS..169..401K} which
contain the off--axis angle and the detected number of source
counts. All our sources have been detected on--axis and the number of
{\it wavdetect}--detected counts is given in Table
~\ref{tab:coor}. This internal positional uncertainty has to be
supplemented with the external uncertainty, which includes the
uncertainty in the satellite aspect solution, and the knowledge of the
geometry and alignment of the spacecraft and focal
plane. \citet{2010ApJS..189...37E} found this external correction to
be 0.39\arcsec, which was subsequently found to be under--estimated by
0.16\arcsec\, by \citet{2011ApJS..192....8R}. The total external 95\%
confidence uncertainty of 0.55\arcsec\, needs to be added in
quadrature to the internal positional uncertainties given in Table
~\ref{tab:coor}.

We use the {\sc ciao} tool {\it specextract} to extract a source
spectrum for each of the three detected sources separately, using the
best known optical coordinates for the sources (see Table~\ref{tab:coor} for references). We created source and
background regions centered on the optical position of the
sources. The circular source regions have a radius of 2\arcsec. The
background regions for \srcone\, and \srctwo\, are annular with inner
and outer radii of 10\arcsec and 30\arcsec, respectively. For \srcthree,
the background is drawn from a
source--free, circular region on the same CCD (because of the smaller sky area covered due to the employment of a
sub--array in the read-out). This circular region has
a radius of 30\arcsec. We do not rebin the extracted source spectra,
although we require each channel to have at least one X-ray photon. We
report the 68\% confidence regions for fitted parameters unless
mentioned otherwise.

\begin{table*}
  \caption{World Coordinate System information of our sample. }
\label{tab:coor}
\begin{center}
\begin{tabular}{cccccccc}
\hline
Source & Optical position & \chan\, X--ray position & 95\% conf.~internal & Total 95\% conf. & Offset & Source & Ref.\\
 &  & &  uncert.~[\arcsec] & uncert. [\arcsec] &[\arcsec] &  counts& $\dagger$ \\ 
\hline
  \srcone & 14:53:13.06 $+$22:14:32.2   &  14:53:13.08
                                         $+$22:14:32.169 & 0.11 & 0.56 & 0.2 & 381 & [1]\\
  \srcone & 223.30442 $+$22.24228 & 223.30449 $+$22.24227
                                                    & 0.11 & 0.56  &0.2 & 381 & [1]\\
\hline  
  \srctwo & 14:57:03.18 $+$49:36:40.97  &  14:57:03.18 $+$49:36:40.865
          & 0.24 & 0.6 &0.1 & 43 & [1]\\
  \srctwo & 224.26325 $+$49.61138 & 224.26326 $+$49.61135&
                                                          0.24      &   0.6   & 0.1 & 43& [1]\\
\hline  
  \srcthree & 11:08:40.12 $+$34:05:52.23 & 11:08:40.13
                                           $+$34:05:53.045 & 0.56 & 0.78 & 0.8 & 8 & [3]\\
  \srcthree & 167.16717 $+$34.09784 & 167.16719
                                        $+$34.09807 & 0.56 &0.78  & 0.8 & 8 & [3]\\
\hline  
  \srcfour & 16:33:55.94 $+$30:14:16.3 & -- & --  & -- & -- & -- & [1]\\
  \srcfour & 248.4831 $+$30.23786 & -- & -- & --  & -- & -- & [1]\\
  
\end{tabular}
\end{center}
\footnotesize{$\dagger$ Reference for the optical coordinates of the sources: [1]~\citet{2014ApJ...793...38A}; [3]~\citet{2014MNRAS.445.3263H}}
\tablecomments{Optical and \chan\, X--ray coordinates of the tidal
    disruption events in our sample, the offset between the two and
    the number of X--ray counts detected in the observation between
    0.3--7 keV. The nominal external uncertainty on the \chan\, X--ray
    coordinates is 0.55\arcsec\, at 95\% confidence. We have chosen to
    add this in quadrature to the provided internal uncertainty in the
    fourth column. For \srcone\, we report the values found in Obs ID
    19532 as this is the longer of the two, providing significantly
    more source counts. The coordinates found when using Obs ID 20879
    are fully consistent with this. }
\end{table*}

We fitted the extracted spectra of each source individually using the
{\sc heasoft} {\sc xspec} tool version 12.10.1. We excluded photons
detected outside the range 0.3--7 keV, as this energy interval is the
best calibrated and most sensitive range for \chan. Throughout the
spectral fitting we employ Cash statistics
(\citealt{1979ApJ...228..939C}) unless mentioned otherwise. For each
source we fit the background spectrum separately first. A power law is
an adequate, first order, description of the background spectrum (see
Table~\ref{tab:xfit}). When fitting the source spectrum, the background
is described using the shape and parameters fixed to those derived
from the separate background fit. We scale the normalization of the 
power law model (that describes the background) on the
basis of the ratio between the size of the source region and that of
the background region.

\subsection{\srcone}

\srcone\, has a redshift of $z=0.1146$ (d$_L=532.6$ Mpc) and is
associated with the galaxy SDSS~J145313.07$+$221432.2
(\citealt{2014ApJ...793...38A}). Given the relatively high observed
count rate of \srcone\, we investigate if the source spectrum is
affected by pile--up by employing the {\sc ciao} tool {\sc
  pileup\_map} on an image created including all photon energies for
both observations of \srcone. The count rate per frame in both
observations is less than 0.02, implying a pile--up fraction lower
than 1\%. Therefore, we conclude that pile--up is insignificant for
our observations of \srcone\, and by extension, given that the other
sources we observed have a lower count rate per frame, those spectra
are not affected by pile--up either.

In the fit we take the attenuating effect of Galactic foreground
extinction into account. To model this effect we use the {\sc xspec}
{\sc phabs} multiplicative model, where we convert the $A_V=0.098$ for
Galactic foreground extinction obtained through NED
(\citealt{2011ApJ...737..103S}) to an $N_H=1.8\times 10^{20}$
cm$^{-2}$ using the relation $N_H=1.79 A_V \times 10^{21}$ cm$^{-2}$
\citep{1995A&A...293..889P}. The
value of $N_H$ is kept fixed during the fit. We employ the {\sc
  xspec} fit--function {\sc pegpwr $+$ phabs $\times$ pegpwr}. For here and below,
we note that in all cases the normalisation of the {\sc pegpwr}
function is equal to the unabsorbed 0.3--7 keV flux.

Fitting the two observations together, the spectrum of \srcone\, is
well--fit by a power law with index $\Gamma$=2.5$\pm$0.1, with an
unabsorbed 0.3--7 keV flux of (9.5$\pm$0.6)$\times 10^{-14}$\flx\,
translating to a 0.3--7 keV luminosity L$_X=(3.2\pm0.2)\times 10^{42}$\lum, where in the calculation of the luminosity uncertainty we, here and below, only included the uncertainty in the flux measurement and not that in the distance determination. The
observed, absorbed, 0.3--7 keV flux is
(8.5$\pm$0.5)$\times 10^{-14}$\flx. The C-statistic of the fit was 226.6 for 223 bins and 221 degrees of freedom. Using the {\sc goodness} command in {\sc xspec} we obtained that 100\% of the realizations yield a lower fit statistic (see \citealt{Jonker2005} for an explanation of what {\sc goodness} values imply). In order to investigate this high value of the {\sc goodness} further, we also fitted the two data sets separately. Whereas the value that we obtain for the power law index is consistent within the 1~$\sigma$ errors, the normalisation is only consistent within 2~$\sigma$. Given the high number of counts detected, we produced a light curve of the observation with ID 19532 with 1 ksec--long bins to investigate if
high amplitude flares are present: none were found, although the source count rate is not constant. Fitting the count rate light curve of 1 ksec--long bins with a constant yields a reduced $\chi^2=2.15$ for 53 degrees of freedom. If spectral variability is associated with this light curve variability this could explain the high {\sc goodness} values of the single observation fit as well as the spectral fit to the data combined. However, there are not enough X--ray photons detected to investigate this further. 

For reference, given the observed number of background events
extracted in the background region (1720 for obs ID 19532 and 479 in
obs ID 20503) one expects that out of the 447 detected counts at the
source position (375 and 72 for the two obs IDs, respectively), 11 are
due to the background (8.5 and 2.4 for the two obs IDs, respectively).

To check our results, we rebinned the data of obs ID 19532 (the longer
of the two observations), requiring that each bin contains at least 30
counts. We subtracted the background and fit the resulting spectrum
with a power law attenuated by the foreground Galactic extinction
employing Chi--squared statistics. The result is fully consistent with
that obtained fitting the unbinned data on both data sets.

\subsection{\srctwo}

\srctwo\, has a redshift of $z=0.064$ ($d_L=287.4$ Mpc) and is
associated with the galaxy SDSS~J145703.17$+$493640.9
(\citealt{2014ApJ...793...38A}).

The spectrum of \srctwo\, is relatively soft compared to the spectrum
of \srcone: no photons with energies above 2 keV are detected. We
fitted the source spectrum with a redshifted black body including a
power law model for the background using Cash statistics. As for
\srcone\, our fit--function includes a factor to model the foreground
extinction, $N_H$. For this we use a rounded--off value of
1$\times 10^{20}$ cm$^{-2}$ given the $A_V=0.046$ from NED
(\citealt{2011ApJ...737..103S}). The value of $N_H$ is kept fixed during the fit.

Fitting the source and background together, we use a fit--function of
an absorbed, redshifted black body for the source plus a power law for
the background ({\sc pegpwr $+$ phabs $\times$ zashift $\times$
  bbodyrad} in {\sc xspec}). We find a best-fit value for the black body temperature of
0.18$\pm$0.02 keV. The unabsorbed source flux, subtracting the flux
due to the background power law in the 0.3--7 keV range is
1.9$\times 10^{-14}$\flx\, giving a 0.3--7 keV luminosity of
L$_X=2 \times 10^{41}$\lum. The absorbed 0.3--7 keV flux is
$(1.7^{+0.3}_{-0.5})\times 10^{-14}$\flx. The C-statistic of the fit was 34.5 for 29 bins and 27 degrees of freedom. Using the {\sc goodness} command in {\sc xspec} we obtained that 98\% of the realizations yield a lower fit statistic (when all simulations are drawn from the best-fit model). The bolometric source luminosity of 2.7$\times 10^{41}$\lum~  is obtained from the normalisation of the black body spectral component. Here we assumed that no other emission components are present in other parts of the spectral energy distribution.

As the fit shows some notable residuals, it mostly under-predicts the flux at low energies, we also try the simple fit--function used for \srcone\, ({\sc pegpwr $+$ phabs $\times$ pegpwr} in {\sc xspec}). For this power law fit we find a best-fit value for the power law index of
3.9$\pm$0.4, and an unabsorbed source flux in the 0.3--7 keV range of 
$3.9^{+1.2}_{-0.9}\times 10^{-14}$\flx\, giving a 0.3--7 keV luminosity of
L$_X= 3.9^{+1.1}_{-1.0}\times 10^{41}$\lum. The absorbed 0.3--7 keV flux is
$(3.5\pm0.9)\times 10^{-14}$\flx. The C-statistic of the fit was 25.2 for 29 bins and 27 degrees of freedom. Using the {\sc goodness} command in {\sc XSPEC} we obtained that 58\% of the realizations yield a lower fit statistic (again when all simulations are drawn from the best-fit model). 

\begin{figure*}[ht]
  \centering
  \includegraphics[width=.48\linewidth]{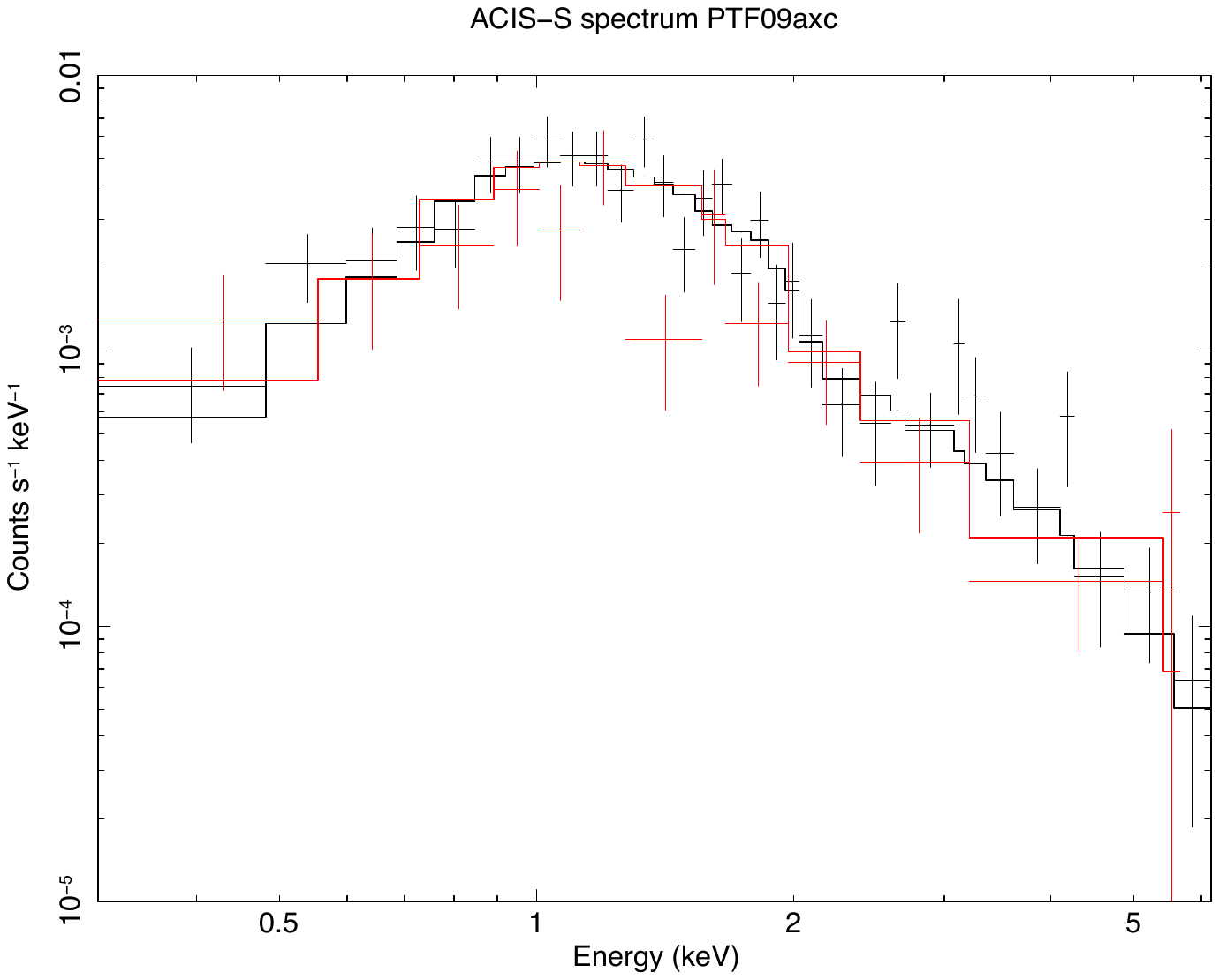}
  \includegraphics[width=.48\linewidth]{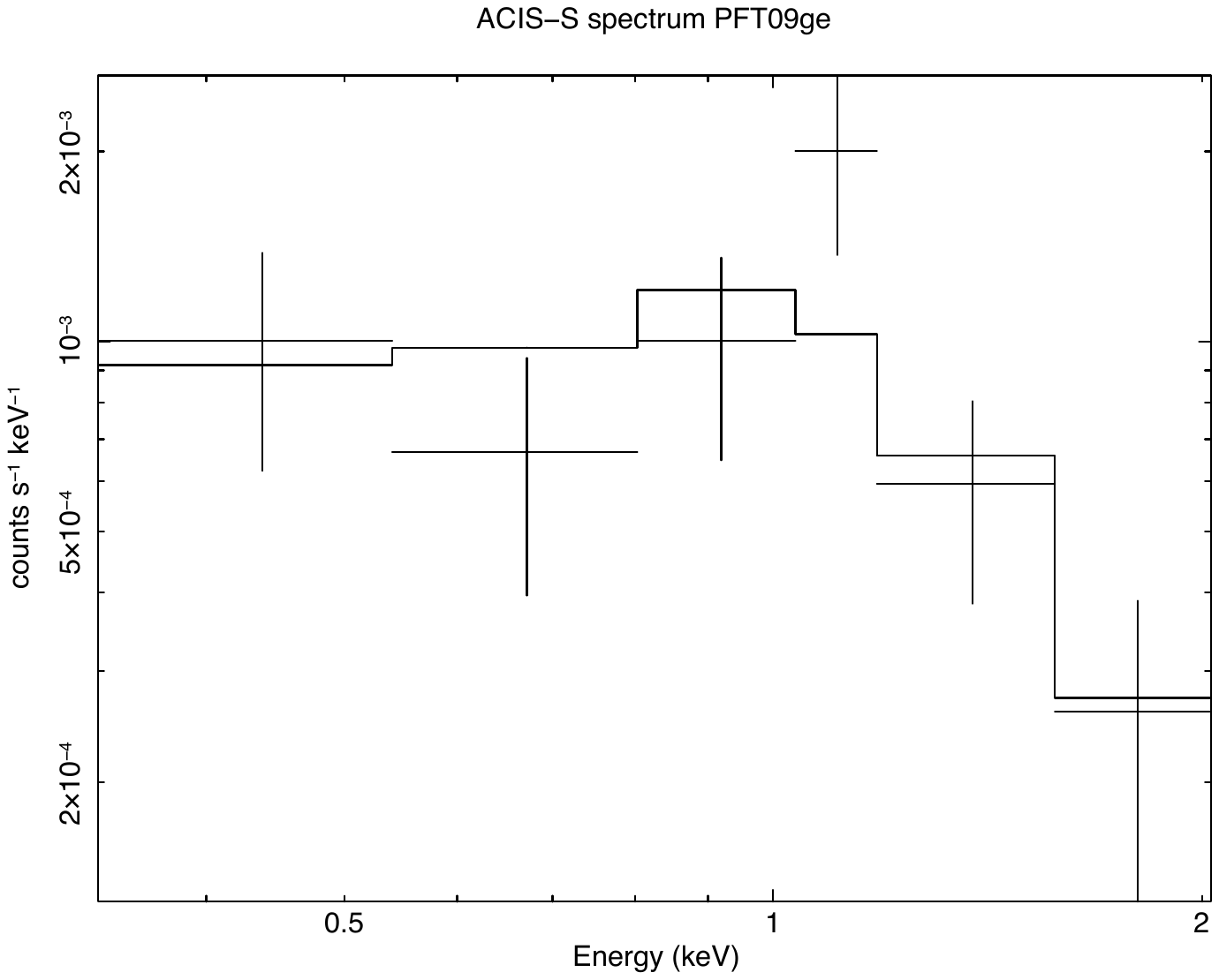}
  \caption{{\it Left panel:} We show the \chan\, ACIS--S spectrum of PTF09axc fitted with a power law folded through the detector response. The black data points are from observation ID 19532 and the red (/grey) data points are from observation ID 20879. The best--fit power law index is 2.5$\pm$0.1. {\it Right panel:} The \chan\, ACIS--S spectrum of PTF09ge fitted with a power law. The best--fit power law index is 3.9$\pm$0.4.}
  \label{plot:xspectra}
\end{figure*}

\subsection{\srcthree}

\srcthree\,has a redshift of $z=0.043671$ ($d_L=193.3$ Mpc) and is
associated with the galaxy SDSS~J110840.11$+$340552.2
(\citealt{2014MNRAS.445.3263H}). For foreground extinction, $N_H$, we
use a rounded--off value of 1$\times 10^{20}$ cm$^{-2}$ given the
$A_V=0.048$ from NED (\citealt{2011ApJ...737..103S}). The
value of N$_H$ is kept fixed during the fit.

Eight photons are detected at a position consistent with that of the
optical source in outburst. Owing to the relative short exposure
compared to the other observations we report on in this manuscript, on average only 0.3
background counts would fall in the source extraction region. Given this very low back ground event rate the eight--count detection is highly significant: i.e.~it occurs due to chance in approximately one out of 8$\times 10^8$ cases. For our spectral analysis of these eight photons we do not correct for this expected background. We fitted for the
power law index and normalisation in the fit--function {\sc phabs
  $\times$ pegpwr} in {\sc xspec}. The best--fit power law index is
$\Gamma=$ 3.2$\pm$1.0. The unabsorbed 0.3-7 keV flux is
$(2^{+2}_{-1})\times 10^{-14}$\flx, giving a 0.3--7 keV luminosity of
$(9^{+9}_{-5})\times 10^{40}$\lum. The absorbed 0.3--7 keV flux is
(1.8$\pm0.8$)$\times 10^{-14}$\flx. The C-statistic of the fit was 17.6 for 8 bins and 6 degrees of freedom. Using the {\sc goodness} command we obtained that 96\% of the realizations yield a lower fit statistic (again when all simulations are drawn from the best-fit model). This high {\sc goodness} shows that the spectral shape is ill--constrained given the low number of detected X--ray photons.

\subsection{\srcfour}

\srcfour{} has a redshift of $z=0.184$ ($d_L=893.2$ Mpc) and is associated
with the galaxy SDSS~J163355.96$+$301416.6
(\citealt{2014ApJ...793...38A}). For foreground extinction, $N_H$, we
use a rounded--off value of 1$\times 10^{20}$ cm$^{-2}$ given the
$A_V=0.049$ from NED (\citealt{2011ApJ...737..103S}). The
value of $N_H$ is kept fixed during the fit.

No X--ray photons with energies between 0.3--7 keV have been
detected in a circle with a radius of 1\arcsec\, centered on the
optical outburst position of \srcfour. We estimate the average
background photon rate in 0.3--7 keV by extracting the detected counts
in a circular region with a radius of 30\arcsec\, close to the source
where no sources were found when using the {\sc wavdetect} tool with
default parameters. 110 background photons are detected in such a region
centered on coordinates RA 16:33:52.17 Dec.~$+$30:13:44.9,
implying that on average 0.12 background count is expected in a
1\arcsec\, circular region.

Following \citet{1991ApJ...374..344K} and
\citet{1984NIMPA.228..120H}, we derive a 95\% confidence upper limit
on the number of detected source counts in the 0.3--7 keV band of 3.
To convert this to a limit on the flux, we divide the upper limit on
the detected number of source counts by the on--source time of this
observation to obtain an upper limit on the source count rate. Next,
we use two models for the spectral shape of the source: a blackbody
with a temperature of 180 eV similar to that found for \srctwo, or a power law
with index of 2.5, as was found for \srcone. The attenuating effect of the $N_H$ derived above is marginal and therefore ignored. W3PIMMS\footnote{https://heasarc.gsfc.nasa.gov/cgi-bin/Tools/w3pimms/w3pimms.pl} provides a
95\% upper limit to the (absorbed) 0.3--7 keV X--ray flux of
F$_X \leq 3 \times 10^{-15}$ \flx\, both for the power law model and
for the black body model. This yields an upper limit to the source 0.3--7 keV 
luminosity of L$_X \leq 3 \times 10^{41}$ \lum\, for both models.

\begin{table*}
  \caption{X--ray spectral fit--parameters for \srcone, \srctwo\, and \srcthree. The normalization of the background has been scaled down to match the source photon extraction area (the scaling factor was 200 for \srcone\, and \srctwo\, and 900/4 for the circular background region for ASASSN--14ae).}
\label{tab:xfit}
\begin{center}
\begin{tabular}{cccccc}
\hline
  Source & Background & Background & Source model & Source flux (absorbed; 0.3--7 keV) & Luminosity (0.3--7 keV)\\
              & Power law index $\Gamma$   & Flux \flx  &  Power law index  $\Gamma$             &   Flux \flx      &  \lum\\
              \hline
  \srcone & 0.7$\pm$0.1 & 3.2$\times 10^{-16}$& 2.5$\pm$0.1 & $(8.5\pm$0.5)$\times 10^{-14}$ & $(3.2\pm0.2)\times 10^{42}$\\
  \srctwo & 0.3$\pm$0.2 & 3.7$\times 10^{-16}$& 3.9$\pm$0.4 & $(3.5\pm0.9)\times 10^{-14}$ & $3.9^{+1.1}_{-1.0}\times 10^{41}$\\
  \srcthree & -- & -- & 3.2$\pm$1.0& $(1.8\pm0.8)\times 10^{-14}$ & $(9^{+9}_{-5})\times 10^{40}$\\
 \srcfour & -- & -- & 2.5$^*$ & $<3\times 10^{-15}$ & $<3\times 10^{41}$ \\ 
\end{tabular}
\end{center}
\footnotesize{$^*$ Parameter fixed to this value. When using a 0.18 keV black body to convert the derived upper limit on the number of source photons to flux as same flux limit as reported for the power law spectral shape is obtained.}
\end{table*}

\section{Discussion}
\label{sec:disc}
We observed four optically--selected TDEs in X--rays using the \chan\, satellite. One source, \srcthree, was observed 4.8 yr after its discovery by \citet{2014MNRAS.445.3263H}, while the other three sources were observed $\approx$8--10 yr after their discovery in 2009 (note that these three TDEs were reported in the literature 5 years after their discovery, by \citealt[][]{2014ApJ...793...38A}). Three of the four sources were detected; only \srcfour\, remains undetected. The X--ray detections of \srcone\, and \srctwo\, 
are especially interesting in conjunction: the X--ray spectrum of \srcone{} is well--fit with
a power--law ($\Gamma=2.5\pm0.1$); conversely, our observations of \srctwo{} are 
well--fit by a very soft, $\Gamma=3.9\pm0.4$, power law. Finally, for \srcthree{}, 
the number of detected X--ray photons is too low for a meaningful spectral fit.  Our {\it Chandra} detections are consistent with the 2014 upper limit of L$_X<2.3\times 10^{42}~$\lum\, for \srctwo{}, the 2014 detection of L$_X = 7.1_{-3.1}^{+12}\times 10^{42}~$\lum\, for \srcone{} \citep{2014ApJ...793...38A}, and the 2014 upper limit of $L_{\rm X} < 1.3 \times 10^{41}~{\rm erg~s}^{-1}$ for \srcthree{} \citep{2014MNRAS.445.3263H}. 

Our results indicate that optically--selected TDEs may maintain a substantial X--ray luminosity for at least $\sim 5-10~{\rm yr}$ post-peak, long after the optical emission has become undetectable. 
Notably, several optically selected TDEs have stringent early--time
X--ray upper limits around or below the luminosities seen in the three sources we detected at late times. For instance, \citet{2012Natur.485..217G} provide a non--detection for the optical/UV selected TDE PS1--10jh, with an upper limit to the 0.2--10 keV X--ray luminosity of $<5.8\times 10^{41}$\lum. \citet{nadia-iptf} report a marginal detection of the TDE iPTF16fnl in stacked observations with a 0.3--10 keV luminosity of $2.4^{+1.9}_{-1.1}\times 10^{39}$\lum, and \citet{hung-2017} did not detect the TDE iPTF16axa down to a 0.3--10 keV luminosity limit of $<3 \times 10^{41}$\lum. As a caveat, we note that these reported upper limits were provided for the 0.2/0.3--10 keV band, whereas in Table~\ref{tab:xfit}, we report 0.3--7 keV luminosities. For spectral shapes with power-law index of 2 (typically assumed for the above cases), these upper limits would be 10--20\% lower when converted to the 0.3--7 keV band. 

The late--time detection of X--ray emission in \srcone{} and \srctwo{} provides further evidence against the alternative hypothesis that most claimed TDE candidates are, in reality, exotic nuclear supernovae \citep{Saxton+18}.  Supernova (SN) explosions are not generally bright in X--ray wavelengths, and even among those that are X--ray bright, none are observed to emit above $\sim 10^{39}~{\rm erg~s}^{-1}$ at times $\gtrsim 10^4$ days post--peak \citep{Dwarkadas&Gruszko12}. This upper limit is far below even the late--time luminosity detected for \srcone, and \srctwo. However, SN2006JD was observed at a luminosity of $(2.77\pm0.45)\times 10^{41}$\lum\, nearly 4.5 years after the supernova was first detected (\citealt{2012ApJ...755..110C}). This luminosity and also the time after the outburst discovery is consistent with those of \srcthree. Therefore, in that case its late-time X--ray luminosity alone is not sufficient evidence for it being a TDE, however, the combined properties (e.g.~such as presented in \citealt{2014MNRAS.445.3263H} and the late--time X--ray luminosity seems incompatible with a Type IIn supernova). Our observations complement late--time FUV detections of six TDE candidates (including \srctwo{}) by \citet{vanVelzen+18}, which also argue against a ``peculiar SNe'' interpretation for those sources.

Our results also constrain the hypothesis that \srcone{} may represent extreme optical variability in a low-luminosity AGN.  This interpretation was first raised in \citet{2014ApJ...793...38A}, who observed a weak [O~III] emission feature with luminosity $L_{\rm [O~III]}= (2.4\pm 0.3)\times 10^{39}~{\rm erg~s}^{-1}$.  This feature is not conclusive evidence of an AGN, and could also be produced by star formation, but in conjunction with the 2014 X--ray detection of the host galaxy, has cast doubt on the TDE status of \srcone{} (see e.g. \citealt{Auchettl+17}).  Our X--ray luminosity measurement strengthens the case that \srcone{} is indeed a bonafide TDE.  Using an empirical relationship between the [O~III] and 3--20 keV luminosities in AGN \citep{2005ApJ...634..161H}, we can estimate the range of [O~III] line luminosities expected if our X--ray detection were of AGN origin (the scatter in this relationship is $\sigma=0.51$ dex, i.e.~a factor $\approx$3.24). Converting our 0.3--7 keV luminosity to the 3--20 keV band using W3PIMMS, \srcone\,has an $L_X$ (3-20 keV) of $8\times 10^{42}$\lum, and therefore the predicted AGN luminosity for the [O~III] line would be
$L_{\rm [O~III]}\approx 5.7\times 10^{40}$\lum, which is a factor $\approx 24$ higher than the actual  $L_{\rm [O~III]}$ measured by \citet{2014ApJ...793...38A}.  The predicted value of $L_{\rm [O~III]}$ is inconsistent with the observed value at the 2.7$\sigma$ level, making a conventional AGN origin for the X-ray and [O~III] luminosity unlikely.

The detected \chan\, luminosities of \srctwo{} and \srcthree{} can be compared with the late--time FUV luminosities reported for those sources by \citet{vanVelzen+18}.  FUV detections of these sources were used to produce disk models and estimates for a range of thermal soft X--ray luminosities; the range of modeled X--ray predictions is particularly sensitive to the dimensionless SMBH spin parameter, $\chi_\bullet$.  
While our detection of \srcthree{} is compatible with the lower end (i.e.~retrograde disk and large $|\chi_\bullet|$) of the predicted range $\log_{10}[L_{\rm X}/({\rm erg}~{\rm s}^{-1})] = 41.7^{+1.3}_{-0.9}$, our detection of \srctwo{} is considerably brighter than the predicted range $\log_{10}[L_{\rm X} /({\rm erg}~{\rm s}^{-1})] = 37.0^{+3.6}_{-2.6}$ (\citealt{vanVelzen+18}, where the fiducial predictions correspond to assuming $\chi_\bullet=0$, and the lower and upper error bars correspond to assuming $\chi_\bullet = -0.9$ and $\chi_\bullet = 0.9$, respectively).  This discrepancy could be reconciled by invoking even larger values of prograde SMBH spin and/or a SMBH mass somewhat smaller than the fiducial prediction of the $M_\bullet - \sigma$ relationship \citep{wevers-masses-ii}.  Unfortunately, \srcone\, was not observed at late times in the FUV.

Interestingly, \srcfour{}, which went undetected in the X--rays (with a 0.3--7 keV upper limit of $3\times 10^{41}$\lum{}), was detected in the FUV at 3$\times 10^{42}$~\lum{}, leading to a predicted X--ray luminosity range $\log_{10}[L_{\rm X} /({\rm erg}~{\rm s}^{-1})] = 41.5^{+1.6}_{-1.1}$.  Our non--detection is compatible with this prediction for any range of retrograde SMBH spin values.  While there are a number of important caveats associated with the late--time X--ray luminosity predictions from \citet{vanVelzen+18}, the strong sensitivity of quasi--thermal X--ray emission to $\chi_\bullet$ in late--time TDE disks underlines the value of multiwavelength, late--time observations for constraining SMBH spin. We will return to this subject in Section~\ref{sec:retro}.

The late--time X--ray detections of the optically--selected TDEs \srcone, \srctwo, and \srcthree~can be compared with late--time detections of X--ray--selected TDEs such as the TDE in IC~3599 reported by \citet{1999A&A...343..775K}. These authors report that this source underwent a spectral change to a hard spectral state and at 6--6.5 years after the X--ray outburst was discovered the luminosity was $\approx 4\times 10^{40}$\lum. Similarly, the TDE in the dwarf galaxy RBS~1032 was found at a luminosity of $10^{41}$\lum~19 years after discovery (\citealt{2014ApJ...792L..29M}). \citet{2004MNRAS.349L...1V} presented \chan\, X--ray observations of five ROSAT--discovered TDE flares (WPVS~007, IC~3599, RX~J1242.6$-$1119, RX~J1624.9$+$7554 and NGC~5905) ranging between $\sim$9--12 years after their discovery. The late-time luminosity of the sources range from 6$\times 10^{39}$--$2\times 10^{41}$\lum. Out of the five sources these authors could only provide a spectral fit for IC~3599 which showed that the source power--law index was still rather soft ($\Gamma\approx 3.6$) even at $\sim$12 years after discovery. Whereas the spectral shape of the other four sources derived from broad band colours was likely somewhat harder, the low number of detected counts precluded more detailed spectroscopic fits. Overall the situation is comparable to the late--time X--ray luminosities and spectral shapes we report for the optically--detected TDEs \srcone, \srctwo, and \srcthree. Apparently, whether or not a TDE is detected in X--rays at early times does not influence the late--time X--ray luminosity. 

\subsection{Disk state changes}

Stellar--mass black holes that accrete from companion stars are visible as X--ray binaries.  The X--ray emission from these disks exhibits a wide variety of spectral properties, or ``states'' (e.g.~\citealt{1989A&A...225...79H}; \citealt{2004ARA&A..42..317F})\footnote{Formally, both timing and spectral properties are necessary for the identification of states (\citealt{1989A&A...225...79H}).  Regrettably, the low number of detected X-ray photons in our late-time TDE observations precludes us from a meaningful X--ray timing study.}.
Two of the most commonly observed states, the high--soft and low--hard state, are characterized by quasi--thermal and power--law spectra, respectively.  Soft states often show sub--dominant power--law X--ray contributions from thermal seed photons up--scattered by an electron corona.  One of the important variables controlling the accretion state of an XRB disk is the dimensionless mass accretion rate $\dot{m} \equiv \dot{M} / \dot{M}_{\rm Edd}$, where $\dot{M}$ is the physical accretion rate and $\dot{M}_{\rm Edd}$ is the Eddington--limited accretion rate.  Because $\dot{M}$ in X--ray binary disks can vary greatly on humanly observable timescales, state changes are often observed, typically following a hysteresis pattern \citep{Maccarone-hyst2003}.  When a source in a high--soft state experiences a persistent decline in $\dot{m}$, it will typically transition to a low-hard state once $\dot{m}$ falls below a threshold value $\sim 0.03$ \citep{Maccarone-hyst2003}.  However, some variation in this transition luminosity (as a fractional Eddington luminosity) has been observed: \citet{Kalemci+13} find a soft--to--hard X--ray state change at an Eddington ratio of $\dot{m} = 0.0030 \pm 0.0041$, and on the extreme end, ~\citet{2019MNRAS.488L.129C} find a recent outburst of the candidate black hole XRB MAXI~J1535-571 in which the soft--to--hard spectral state change seems to occur at a fraction 1.2--3.3$\times 10^{-5}$ of the Eddington luminosity (see also \citealt{2003A&A...409..697M} for a discussion of variation in Eddington fraction for state changes in XRBs).

There is some evidence that analogous state changes occur in AGN accretion disks around SMBHs (e.g.~\citealt[][and references therein]{maccarone03}). However, as the viscous times in AGN disks are typically much longer than reasonable observational baselines, it is not easy to observe state changes in AGN.  A further difficulty is that in the soft X--rays, AGN spectra are generally dominated by power--law or reflection contributions.  This is because the peak of the thermal blackbody disk emission occurs in the extreme UV, where observations are hindered by gas and dust extinction (although a soft spectral component can sometimes be discerned, e.g. \citealt{done2012}).

Compared to standard AGN, TDE disks are probably more favorable laboratories for observing ``scaled up'' state changes around SMBHs \citep{Giannios&Metzger11, Tchekhovskoy+14}. The main reason is that the accretion disks expected to form in TDEs are much smaller than AGN disks, implying shorter time scales.  If we consider a steady--state Shakura--Sunyaev disk with dimensionless viscosity $\alpha$, constant aspect ratio $H/R$, and an outer edge $R_{\rm d}$, the viscous time scales as $\propto R_{\rm d}^{3/2}$. Late--time TDE disks should be geometrically thin and mostly circularized, and have an outer radius $R_{\rm d} \sim 2R_{\rm p} = 2R_{\rm t}/\beta$, where $\beta\sim 1$ is the penetration parameter of the TDE, and the tidal radius is 
\begin{align}
R_{\rm t} =& R_\star \left(\frac{M_\bullet}{M_\star}\right)^{1/3} \\
\approx & 2 \times 10^{-6}~{\rm pc} \left( \frac{M_\bullet}{10^6 M_\odot} \right)^{1/3} \left( \frac{M_\star}{M_\odot} \right)^{-1/3} \left( \frac{R_\star}{R_\odot} \right). \notag
\end{align} 
Here $M_\star$ and $R_\star$ are the mass and radius of the victim star, and we see that both $R_{\rm t}$ and $R_{\rm d}$ are far smaller than the typical radius of an AGN accretion disk: for example,  if we take the extent of an AGN broad line region as a first order approximation of the size of the AGN accretion disk (e.g.~\citealt{2011A&A...525L...8C}) using the relation that the 5100$\AA$ luminosity $\lambda L_\lambda$ implies a typical scale $R_{\rm BLR} \approx 0.026 \big( \frac{\lambda L_\lambda(5100\AA)}{10^{44}}\big)^{0.7}$ pc \citep{kaspi-2000}, a factor of 10$^4$ times larger than the typical TDE disk.

Shortly after disruption, the peak mass fallback rate onto the SMBH will generally be super--Eddington, with a peak fallback rate $\dot{M}_{\rm peak} = \frac{1}{3}M_\star / t_{\rm fall}$, where 
\begin{equation}
    t_{\rm fall} \approx 3.5\times10^6~{\rm s}~ \left( \frac{M_\bullet}{10^6 M_\odot} \right)^{1/2} \left( \frac{M_\star}{M_\odot} \right)^{-1} \left( \frac{R_\star}{R_\odot} \right)^{3/2} \label{eq:tFall}
\end{equation}
is the fallback time for the most tightly bound debris.  In Eddington units, this is \citep{stone+2013}
\begin{equation}
    \frac{\dot{M}_{\rm peak}}{\dot{M}_{\rm Edd}} \approx 130 \left( \frac{M_\bullet}{10^6 M_\odot} \right)^{-3/2} \left( \frac{M_\star}{M_\odot} \right)^{2} \left( \frac{R_\star}{R_\odot} \right)^{-3/2}.
\end{equation}
If circularization is efficient, the disk accretion rate $\dot{M}$ will track the (super--Eddington) mass fallback rate, and therefore the most relevant stellar--mass point of comparison might seem to be ultra--luminous X--ray sources (ULXs), rather than high--soft XRBs (which are generally sub-Eddington). Contrary to this supposition, early--time soft X--ray detections of TDE candidates generally find quasi--thermal spectra that {\it are} analogous to a high--soft state \citep{1999A&A...349L..45K, Greiner+00}, particularly in TDEs with
good--quality early--time X--ray spectra (e.g.~\citealt{2012A&A...541A.106S, 2015Natur.526..542M, 2015ApJ...811...43L, 2016MNRAS.455.2918H, gezari-15oi-2017, Wevers+19}), although we note that given the limited pass--band (typically 0.2--10 keV at best) and especially in cases where the number of detected X--ray photons is below $\sim$10,000 it is difficult to rule out the soft ULX state. For high signal--to--noise spectra the main difference found by \citet{gladstone2009} is that the ULX state is characterised by a high Componisation optical depth ($\tau \sim$5--20) whereas in the XRB soft state when fitted with the same model $\tau$ is always $\leq$1.

However, even in the limiting case of rapid circularization, the super--Eddington phase is expected to last only a fraction of the time TDEs are typically observed.  Given the absence of observed state changes from a super--Eddington, ULX--like state to a sub--Eddington, high--soft state, we deem it likely that X--ray bright TDEs are seen mostly in the equivalent of the XRB soft state. As we will discuss in Section~\ref{sec:XvsO}, the absence of super--Eddington emission may be related to a delay before the sources are detected in X--rays.  A soft, quasi-thermal spectrum will no longer be a reasonable expectation (i) at late enough times, once $\dot{m}$ becomes very sub--Eddington, 
or (ii) if circularization is highly inefficient and $\dot{m} \ll 1$ always. Because $\dot{M}/\dot{M}_{\rm Edd}$ steadily decreases during late stages of a TDE
flare, we may expect a late--time transition to the SMBH equivalent of the XRB low--hard state.  

Observationally, TDE candidates with soft spectra containing an additional hard, power--law X--ray spectral
components do exist \citep[e.g.~][]{2016MNRAS.463.3813H, saxton2017}, much like XRB soft states where a sub--dominant power-law component
also exists.  Another example is the X--ray selected TDE 2XMMi~J184725.1$-$631724
(\citealt{2011ApJ...738...52L}). It showed an X--ray spectrum that was
well--fit by a soft thermal component with a temperature of
approximately 60 eV plus a (soft) power law with a photon index of
around 3--4 contributing around 10--15\% to the total 0.2--10 keV
luminosity (at the first detection of the outburst, in Sept 2006). The
temperature of the soft component had risen to around 90 eV nine
months later as measured by \xmm{}, with a power--law contribution of
5--10\%. The X--ray spectrum in the TDE candidate RX~J1242$-$1119 changed from a power--law with $\Gamma\approx 5$ (so a very soft spectrum that could also be fit with a blackbody with a temperature of 0.06 keV) to $\Gamma\approx2.5$ at late--times (\citealt{1999A&A...349L..45K,2004ApJ...603L..17K}), signifying a potential state change. Finally, the X--ray spectrum of the TDE AT~2018fyk is well--fit when a power law is added to the fit function. This power law component constitues $\approx30$\% of the unabsorbed flux \citep{Wevers+19}.

Thus, well--studied TDE X--ray spectra are qualitatively closer to an XRB high--soft state than they are to AGN power laws.  The reasons for this are unclear, but likely involve the higher blackbody temperature of TDE disks near the ISCO, due to (i) the smaller SMBH masses in TDEs relative to most AGN \citep[compare the SMBH mass distributions in][]{Woo-Urry-2002,2017MNRAS.471.1694W,wevers-masses-ii}; (ii) the higher early--time Eddington fraction expected for TDEs in comparison to typical AGN \citep{Kauffmann&Heckman09}; (iii) a bias towards prograde spinning SMBHs for X--ray selected TDEs (see $\S$~\ref{sec:rates}) enabling a smaller value for the innermost stable circular orbit (ISCO).  Early--time TDE X--ray spectra often appear even more thermally dominated than the typical XRB high--soft state, possibly indicating difficulty in forming a Compton scattering corona.

Our interpretation of the spectral properties of \srcone{} and \srctwo{} follows straightforwardly from the XRB analogy: \srcone{} has undergone a state change to the SMBH analogue of the low--hard state, but this type of change has not yet occurred for \srctwo{}, which likely remains in an analogue of the high--soft state. This hypothesis is complicated by the Eddington ratios we observe.  Using literature estimates for the SMBH masses \citep{wevers-masses-ii} and accounting for both the one--sigma scatter of the underlying $M_\bullet - \sigma$ relation and the uncertainty in our X-ray luminosity estimates, we find that \srcone{} was observed at an Eddington fraction of $\dot{m} = 5.4_{-3.8}^{+12}\times 10^{-2}$; \srctwo\, was observed at an Eddington fraction of $\dot{m} = 1.6_{-1.1}^{+3.3}\times 10^{-3}$; and \srcthree{} at $\dot{m} = 2.8_{-2.4}^{+13}\times 10^{-3}$.  The simplest theoretical expectation might be that the TDE disk with the lower Eddington ratio, \srctwo{}, should have undergone a state change prior to one with a higher Eddington ratio (\srcone{}).  However, we note that in XRBs, the emergence of a coronal
power--law and the ensuing state change is regulated not only by the accretion rate $\dot{m}$ but also by an additional parameter (cf.~\citealt{2001ApJS..132..377H}, where the second parameter is interpreted as the fractional size of the Comptonizing region).  Furthermore, TDEs differ from standard accretion disks in several ways, and there are other plausible ``hidden
variables'' that may be acting to prevent the emergence of a corona in
\srctwo.  For example, the relatively weak magnetic fields of main
sequence stars may mean that TDE disks are born with extremely low
magnetizations\footnote{Indeed, TDE disks may be so starved of
  magnetic flux that initial angular momentum transport may be
  dominated by exotic processes such as the Papaloizou-Pringle
  instability \citep{2018MNRAS.474.1737N} or fallback shocks \citep{Chan+19} rather than the usual
  magnetorotational instability.}. Since coronal electron populations
are thought to be accelerated to relativistic energies in magnetic
reconnection events \citep{2001MNRAS.321..549M}, standard low--hard state
coronae may only emerge in TDE disks born with unusually large
magnetizations, or ones where external factors like large and retrograde
SMBH spin \citep{Parfrey2015} favor magnetic field generation {\it in situ} through dynamo processes.  

Overall, the X--ray Eddington ratio of \srcone{} is broadly compatible with the common range of Eddington ratios where soft--to--hard state changes occur in XRBs.  The persistently soft spectrum of \srctwo{} is more unusual, but as mentioned before, XRB soft states have been observed to persist down to an Eddington ratio of $\sim 10^{-3}$ and in an extreme case even down to a few times $10^{-5}$. As an alternative, the different X-ray spectral shape between sources and in some cases the changing X-ray spectral slope in one source can also be reproduced by changing parameters involving a different physical model. One possibility includes varying the Comptonization optical depth and/or the disk  temperatures of the seed photons of such a Comptonization model. 

One testable prediction of our XRB analogy is the predicted radio luminosity using the Fundamental Plane of black hole activity (\citealt{2003MNRAS.345.1057M}; \citealt{2004A&A...414..895F}). Using the calibration of \citet{2003MNRAS.345.1057M}, and given the SMBH mass estimate of $\log {M}_\bullet=5.68$ in \srcone\, from \citet{2017MNRAS.471.1694W}, we derive an expected radio luminosity at 5 GHz of $2\times 10^{37}$\lum. Given the luminosity distance of \srcone{}, this translates to a flux density at 5 GHz of 20 $\mu$Jy, a level which is detectable with current radio telescopes, although this flux estimate carries a substantial uncertainty.

If the soft X--ray spectra of X--ray bright TDEs imply that those systems accrete in the equivalent of the XRB soft state, the fact that many TDEs have very weak or nonexistent early-time radio emission is unsurprising (cf.~\citealt{maccarone03}; \citealt{2013A&A...552A...5V}). We note that XMMSL1~J074008.2$-$853927, another TDE with an X--ray power--law component (index $\Gamma=2$) was detected in radio (\citealt{2017A&A...598A..29S, Alexander+17}), although XMMSL2~J144605.0$+$685735, which shows a power--law with index $\Gamma=2.5$, was not (\citealt{2019A&A...630A..98S}).

Finally, we note that our X--ray detections demonstrate that late--time TDE disks do not generally exhibit a different sort of state change: a collapse into a cold, gas pressure--dominated state due to the development of a thermal instability.  This type of collapse is predicted by simple applications of the popular $\alpha$--disk model, but would imply that late--time TDE disks have luminosities far below what we observe \citep{Shen&Matzner14}.  Our observations further substantiate this point, which was recently made in the context of late--time detections of TDE disks in the FUV \citep{vanVelzen+18}.  The evidence against very cold disks in (most) TDEs seen at late times could indicate that the nonlinear development of the thermal instability is suppressed by an iron opacity bump \citep{Jiang+16}, or alternatively magnetic pressure support \citep{Begelman&Pringle07, Sadowski16, Jiang+19}.

\subsection{Optical vs.~X--ray selected TDEs}
\label{sec:XvsO}

Many of the first TDE candidates were detected from their soft X--ray emission, but either lacked contemporaneous searches for optical variability \citep{1999A&A...349L..45K}, or were observed {\it not} to show variable optical behavior (\citealt{Greiner+00}; \citealt{2012A&A...541A.106S}; \citealt{2019A&A...630A..98S}).  Later, optical and UV surveys discovered a second class of TDE candidates, which often possessed upper limits on their X--ray emission (\citealt{2012Natur.485..217G}; see also \srctwo{}, \srcthree{}, and \srcfour{}).  More recently, a number of TDEs have been observed to exhibit both optical/UV {\it and} X--ray variability \citep[e.g.][]{2016MNRAS.455.2918H, 2016MNRAS.463.3813H, Wevers+19}.  With such a diversity of X--ray ($L_{\rm X}$) and optical ($L_{\rm opt}$) luminosities, it is fair to ask: do these transients all really stem from the same underlying type of event?

In the context of the reprocessing paradigm, this question has sometimes been answered (theoretically) in the affirmative by introducing a viewing angle dependence, akin to the AGN unification model \citep{Metzger&Stone16, Jane-unifi2018, Lu&Bonnerot19}: edge--on TDEs obscure the X--rays from the inner accretion flow, but face--on TDEs are viewed through a low--density polar region, and thus will be X--ray bright. Some observational characteristics such as the emission lines in the optical part of the spectrum becoming more narrow as the TDE evolves, are consistent with this interpretation \citep{leloudas+19}.  The complicated three--dimensional geometry of the circularization/shock paradigm \citep{Piran+15, Shiokawa+15} likely suggests a viewing angle dependence as well.
 
A different -- possibly complementary -- way to unify TDE candidates across a broad range of $L_{\rm X}/L_{\rm opt}$ ratios is to postulate a strong temporal dependence in $L_{\rm X}/L_{\rm opt}$.  Our late--time detections of \srcone{}, \srctwo{}, and \srcthree{} demonstrate that a substantial fraction of optically selected TDEs are X--ray bright at late times $\approx 5-10~{\rm yr}$ post--peak, signifying the presence of an exposed, compact accretion disk.  If the optical emission is caused by circularization shocks, a delay between optical and X--ray would be related to delays in forming the (inner, X--ray emitting) accretion disk, as has been suggested by \citet{Shiokawa+15}. If the optical is instead caused by reprocessing of the inner disk's X--rays and EUV, then an enshrouded inner disk will only become visible in X--rays after the reprocessing screen has diluted enough to permit an ionization breakout \citep{Metzger&Stone16, roth+2016}.

Because the low $L_{\rm X}$ values we observe are compatible with past X--ray non--detections (or, in the case of \srcone{}, its 2014 detection), we are unable to say whether this truly represents {\it brightening} of initially X--ray dim TDEs. Deep limits on the X--ray luminosity in several other optically selected TDEs suggests that brightening is certainly plausible (for references and limits see the first paragraphs of the Discussion).

The nature of the X--ray light curve in optically selected TDEs is a crucial observable to constrain with future observations.  The offset between the peaks of optical and X--ray emission, $\Delta t_{o-X}$, is a key parameter for testing the idea of unification in {\it time}, rather than (or in addition to) angle.  The distributions of $\Delta t_{o-X}$ will depend on the emission mechanism for the optical and X--ray light, as well as on event parameters such as $\beta$, $M_\bullet$, and $\chi_\bullet$. 

Depending on the delay between disruption and X--ray observation, an individual TDE could be in the equivalent of the soft X--ray spectral
state, or as in the case of XMMSL2~J144605.0$+$685735, in a hard power--law like spectrum\footnote{A potential selection effect might be at play, as massive brightening of an X-ray power law is more difficult to separate from AGN flares, and thus will often not be classified as a TDE, but as an AGN flare}.  We hypothesize that the X--ray selected TDEs are, in this scenario, often discovered much longer after the disruption than are optically selected TDEs.  This particular unification hypothesis would be falsified if observations months to years before the X--ray turn--on in a TDE candidate did not show signs of an optical enhancement\footnote{In individual host galaxies, there could be reasons why the optical emission should be strongly reduced in these TDEs (such as the presence of a large amount of nuclear dust, e.g.~\citealt{2018Sci...361..482M}).}. 

This scenario also implies that all optically selected TDEs will at some point emit X--ray radiation, as is true for three of the four sources we observed in this work. Sources which are detected in both optical and X--ray observations at early times (e.g.~ASASSN--14li, ASASSN--15oi and AT2018fyk; \citealt{2016MNRAS.455.2918H}, \citealt{2016MNRAS.463.3813H}, \citealt{Wevers+19}, respectively) could be explained in this scenario as sources with efficient circularization due, for instance, to high $\beta$, large $M_\bullet$ (though this is disfavored by $M_\bullet-\sigma$ BH mass estimates), or large and retrograde SMBH spin.

The shape of the X--ray spectra as well as the lower luminosities that we observed in \srcone, \srctwo\, and \srcthree\, differ from the (soft) X--ray discovered TDEs, which have soft thermal spectra and luminosities of order L$_X\approx 10^{43-44}$\lum\,(\citealt{Auchettl+17}). This implies that our observed sources are, in this scenario, at an even later stage in the evolution of the mass fall--back and accretion rate.

\subsection{Rates of detection in future X--ray surveys}
\label{sec:rates}
Near--future wide--field X--ray surveys are predicted to expand our sample of
X--ray TDEs by $1-2$ orders of magnitude.  For example, the Einstein Probe is
expected to find $\sim 100$ new TDEs per year \citep{2015arXiv150607735Y},
while eROSITA is expected to find $\sim 1000$ \citep{2014MNRAS.437..327K}.
In this section, we revisit the latter estimate, making the following
modifications to the model of \citet{2014MNRAS.437..327K}:

\begin{enumerate}
    \item We allow (in one of our models) the temperature at the inner edge of the accretion disk to be a function of SMBH spin.
    \item We assume the volumetric TDE rate is given by:
    \begin{equation}
        \dot{N}_{\rm tde}=2.9\times 10^{-5} \left(\frac{M_\bullet}{10^8 M_{\odot}}\right)^{-0.4} {\rm yr}^{-1} \phi(M_\bullet).
        \label{eq:tdeRate}
    \end{equation}
    This assumption takes
	a theoretical (per galaxy) TDE rate calibrated from observations of nearby galactic nuclei \citep{stone&metzger2016}, and multiplies this by $\phi(M_\bullet)$, the
	$z=0.02$ black hole mass function from \citet{shankar+2009} (their table 3).\footnote{eROSITA would be sensitive to TDEs with $z\lesssim$0.2, and the \citet{shankar+2009} mass function varies little in this redshift range.} We consider black hole masses between $10^5 M_{\odot}$ and $10^8 M_{\odot}$ in
	our estimate \footnote{Most of the TDEs are expected from the lower end in this mass range. Indeed, \citealt{2017MNRAS.471.1694W} and \citealt{wevers-masses-ii} show that using and extra-polating the M--$\sigma$ correlation, the masses of known TDEs favor the low--mass range}. The volumetric TDE rate is $\sim 10^{-5}$ Mpc$^{-3}$ yr$^{-1}$
	for this range.
        \item{We require the detection of at least 40 counts, at a location where previously no source was detected by eROSITA.}
\end{enumerate} 
    
    We consider two different models for the TDE light curve and spectrum: (I) an optimistic theoretical model based on simple accretion disk theory and (II)
    a more pessimistic quasi--empirical model that is calibrated to reproduce the late--time X--ray properties of \srctwo.  In both cases, we only consider disruption of Solar--type stars, for simplicity. The rates derived below are the X--ray detection rates, with no predictions about the fraction that is optically bright (see Section~\ref{sec:XvsO}).
    
    \subsubsection{Model I}
    \label{sec:theoryRates}
    We assume circularization occurs efficiently, and that the mass accretion rate through the disk is
    \begin{align}
        &\dot{M}_{\rm acc} (M_\bullet, t)=
        \begin{cases}
            0 & t< t_{\rm fall}\\
            \dot{M}_{\rm max}(M_\bullet) \left[\frac{t}{t_{\rm fall} (M_\bullet)}\right]^{-1.2} & t\geq t_{\rm fall}
        \end{cases}
        \label{eq:lbol}
    \end{align}
   where $t_{\rm fall}$ is the fallback time (Eq.~\ref{eq:tFall}).  
   This power law is shallower than the canonical $t^{-5/3}$ decline of the mass fall--back rate and is motivated by theoretical models for viscously spreading disks \citep{Cannizzo+90}, the late time FUV light curves of TDEs \citep{vanVelzen+18}, and our own
   late--time X--ray detections. The maximum accretion rate $\dot{M}_{\rm max}$ is a factor of
   $\sim$3 smaller than the peak fall--back rate $\dot{M}_{\rm peak}$. With this normalization, a total of half a solar mass of material is accreted. 
   
   The bolometric disk luminosity after one fallback time is 
    \begin{align}
        &L_{\rm bol}(t, M_\bullet, \chi_\bullet)=\min[L_{\rm Edd}(M_\bullet), \eta_\bullet(\chi_\bullet) \dot{M}_{\rm acc}(t) c^2]\nonumber\\
        &=\min\Bigl[L_{\rm Edd}(M_\bullet),\nonumber\\ &3\times 10^{45} \left(\frac{\eta_\bullet(\chi_\bullet)}{0.057}\right)  \left(\frac{M_\bullet}{10^6 M_{\odot}}\right)^{-1/2} \left(\frac{t}{t_{\rm fall}}\right)^{-1.2} \mathrm{erg\,\, s^{-1}}\Bigr],
        \label{eq:lpeak}
    \end{align}
    where $L_{\rm Edd} (M_\bullet)$ is the Eddington luminosity, and $
    \eta_\bullet$ is the standard radiative efficiency of a thin, equatorial accretion disk\footnote{The efficiency $\eta_\bullet$ ranges from $0.038$ to $0.42$ as $a_\bullet$ goes from -1 to 1, and is computed as in \citet{Bardeen+72}.}. Here we have further assumed that the disk aligns itself into the SMBH equatorial plane after an initial period of misalignment.  Typical alignment timescales are $\lesssim 100~{\rm d}$ for large ($\chi_\bullet >0.5$) SMBH spins \citep{Franchini+16}, so alignment is a reasonable approximation for eROSITA observations, which have a typical cadence of 6 months\footnote{In principle, alignment can take longer than 6 months if $\chi_\bullet \lesssim 0.5$, but $\eta_\bullet$ is considerably less sensitive to SMBH spin in this regime.}.  Eq.~\eqref{eq:lpeak} is close to the estimated bolometric luminosity of \srctwo\, near peak ($\sim 8\times 10^{44}$ erg s$^{-1}$, which is comparable to the Eddington limit for this source; see \citealt{vanVelzen+18}). 

    Equations~\eqref{eq:lbol} and~\eqref{eq:lpeak} specify the bolometric
    luminosity, but here we are interested in soft X--ray observations of TDEs, and many optically selected TDEs (including three sources of our sample) have not been detected in X--rays at early times. Theoretically, TDEs may become X--ray bright when the central engine
    ionizes through a surrounding reprocessing layer
    \citep{Metzger&Stone16, roth+2016} or, if circularization is inefficient, after repeated
    shock interactions near stream apocenter (e.g.~\citealt{Dai+15, Shiokawa+15}). The
    precise time when this occurs is uncertain. However, at least the early,
    super--Eddington phases of mass fallback are likely to be X--ray
    dim.\footnote{At least for most viewing angles: observers aligned with the
    poles may see X--ray emission from a jet according to the unification model
    of \citealt{Jane-unifi2018}.}  If disk formation is inefficient, there is little accretion to produce X--rays \citep{Shiokawa+15}; even if disk formation is efficient, the inner disk can be heavily obscured by bound debris \citep{Loeb&Ulmer97, coughlin&begelman2014} or by outflows \citep{Miller15, Metzger&Stone16, Jane-unifi2018, Lu&Bonnerot19}. However, as our present work shows, a large fraction of TDEs become X--ray bright at later times, when the luminosity becomes sub--Eddington. This occurs after  
    \begin{align}
        t_{\rm Edd} (M_{\bullet}, \chi_\bullet)\approx 1.5 {\rm yr} \left(\frac{M_\bullet}{10^6 M_\odot}\right)^{-3/4} \left(\frac{\eta(\chi_\bullet)}{0.057}\right)^{5/6}.
    \end{align}
    Here, $t_{\rm Edd}$ is the time after which the accretion rate becomes sub--Eddington.
    In practice, we consider a TDE at redshift $z$ to be detectable by eROSITA after $t_{\rm
    Edd} (M_{\bullet}, \chi_\bullet)$, as long as 
    \begin{align}
    	&\frac{L}{4 \pi d^2_L(z) K(z)}\geq f_{\rm lim},
    \end{align}
    where $d_L(z)$ is the luminosity distance and
    \begin{align}
		&f_{\rm lim}= \frac{C_{\rm crit}}{t_{\rm int} \int_{\nu_{\rm min}}^{\nu_{\rm max}}  \frac{S_{\nu} (\nu) A(\nu) e^{-\xi(\nu)}}{h \nu} {\rm d} \nu}\nonumber\\
		&K(z)^{-1} = \frac{(1+z) \int_{\nu_{\rm min}}^{\nu_{\rm max}} \frac{S_\nu (\nu (1+z)) A(\nu) e^{-\xi(\nu)}}{h \nu}  {\rm d}\nu}  {\int_{\nu_{\rm min}}^{\nu_{\rm max}} \frac{S_\nu (\nu) A(\nu) e^{-\xi(\nu)}}{h \nu}  {\rm d}\nu}
    \end{align}
    Here $C_{\rm crit}$ and $t_{\rm int}$ are the minimum number of counts
    resulting in a detection and the integration time respectively (which we take to be 40 and 240
    seconds following \citealt{2014MNRAS.437..327K}), $A_{\nu}$ is the effective area as a function of energy\footnote{\url{https://wiki.mpe.mpg.de/eRosita/erocalib\_calibration}},
    $e^{-\xi(\nu)}$ accounts for photoelectric absorption\footnote{This is derived 
    from the XSPEC PHABS multiplicative model with $N_H=5\times 10^{20}$ cm$^{-2}$ following \citet{2014MNRAS.437..327K}.},  and $S_{\nu}$ is
    the Spectral Energy Distribution (SED), which we take to be a black--body with an effective temperature corresponding to the temperature near the ISCO as given by Eq.~9 of \citet{Lodato&Rossi11}.\footnote{The effective temperature actually goes to zero at the ISCO in this model. In practice we evaluate $T_{\rm eff,in}$ at 1.36 times the ISCO, where the effective temperature is maximized.}  We integrate SEDs between $h\nu_{\rm min} = 0.2~{\rm keV}$ and $h\nu_{\rm max} = 2~{\rm keV}$ (following \citealt{2014MNRAS.437..327K}).

    The total number of new TDEs detected every year is 
    \begin{align}
        N_{\rm det}=&\int_{0}^{1\,\, {\rm year}} \int_{M_{\rm min}}^{M_{\rm max}} \int_{0}^{z_{\rm lim}(M_\bullet)} \frac{{\rm d}N}{{\rm d}t {\rm d}M_\bullet {\rm d}z} {\rm d}z {\rm d}M_\bullet {\rm d}t\nonumber\\
        =&\int_{0}^{1\,\, {\rm year}} \int_{M_{\rm min}}^{M_{\rm max}} \int_{0}^{V_{\rm c}[z_{\rm lim}(M_\bullet)]} \dot{N}_{\rm tde} {\rm d} V_{\rm c}(z) {\rm d}M_\bullet {\rm d}t,
    \end{align}
    where ${\rm d}N/{\rm d}t {\rm d}M_\bullet {\rm d}z$ is the differential TDE rate per unit SMBH mass per unit redshift, and $z_{\rm lim}$ is the maximum redshift to which a TDE in a given mass bin could be detected. In the second line, $\dot{N}_{\rm tde}$ is the volumetric TDE rate (Eq.~\ref{eq:tdeRate}), while ${\rm d}V_c$ is the co--moving volume element. Conservatively, $z_{\rm lim}$ satisfies
    \begin{align}
        \frac{L(t_o+6 \,\,{\rm months})}{4 \pi d_L(z_{\rm lim})^2 K(z_{\rm lim})}=f_{\rm lim}\nonumber\\
        t_o=\max[t_{\rm edd}(M_\bullet, \chi_\bullet), t_{\rm fall}],
        \label{eq:zlim}
    \end{align}
    where $t_o$ is when the X--rays turn on and six months is the time it takes eROSITA to scan the entire sky. 

\begin{figure}
    \centering
    \includegraphics[width=8.5cm]{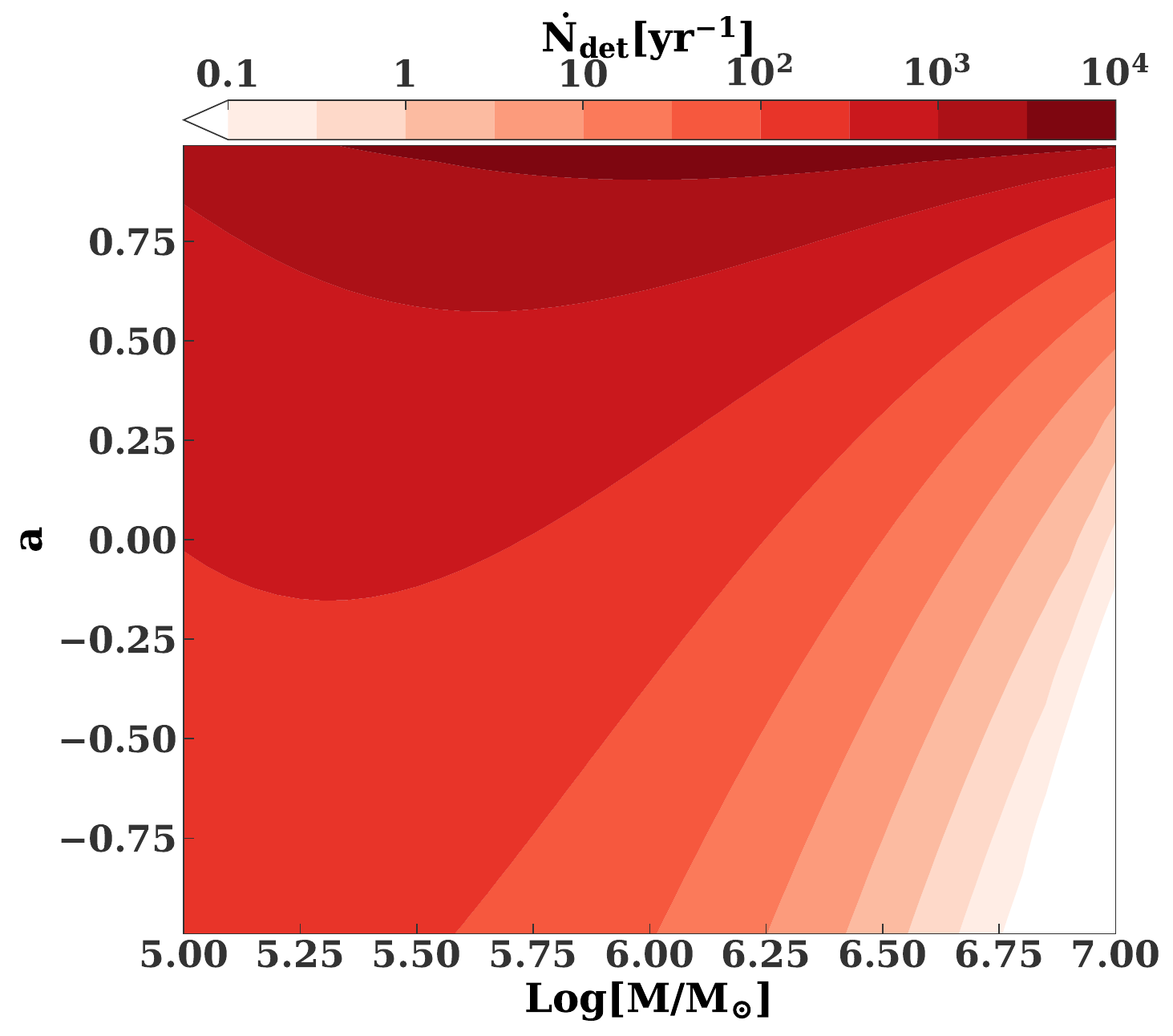}
    \includegraphics[width=8.5cm]{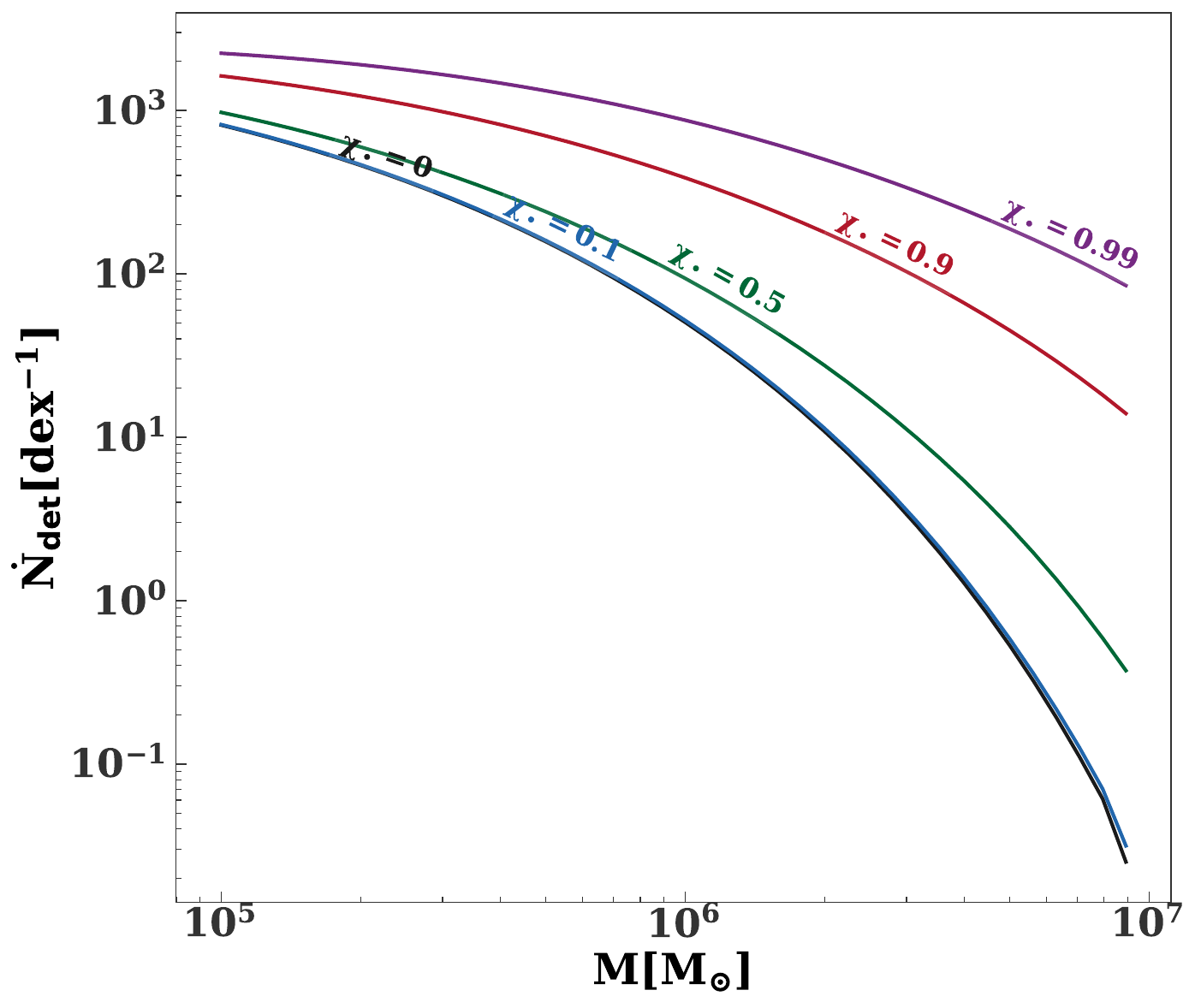}
    \caption{{\emph{Top panel:}} The rate of eROSITA
    detections as a function of SMBH mass and spin, 
    for a fixed TDE rate of $10^{-5}$ Mpc$^{-3}$ yr$^{-1}$. These are the detection rates assuming 
    TDEs are distributed as a delta function with a specific mass 
    and spin (i.e.~each SMBH mass and spin is assumed equally likely). {\emph{Bottom panel:}} The rate of eROSITA detections as a function 
    of SMBH mass, for a selection of spin parameters. These lines are a convolution 
    of the rates from the top panel with the SMBH mass function and a theoretical estimate of TDE rates as a function of $M_\bullet$ (Eq.~\ref{eq:tdeRate}). We assume $50\%$ of TDE disks align into prograde equatorial and $50\%$ align into retrograde equatorial configurations by the time of observation--see the discussion in $\S$~\ref{sec:retro}).
    }
    \label{fig:detRate}
\end{figure}

    The top panel of Fig.~\ref{fig:detRate} shows the eROSITA detection rate
    as a function of SMBH mass and spin, assuming all TDE hosts have the same mass and spin combination, and that the total TDE rate is $10^{-5}$
    Mpc$^{-3}$ yr$^{-1}$. For a flux--limited sample of TDEs produced by rapidly spinning black holes, there are 1--2 orders of magnitude more detections when the black hole spin is universally prograde (with respect to the accretion disk's rotation) than universally retrograde, irrespective of the SMBH mass bin we consider. In stark contrast to optically selected TDE samples ($\S$~\ref{sec:retro}), an X--ray selected sample would be strongly biased towards prograde black hole spins, though this bias abates if the SMBH spin distribution is very bottom--heavy (with typical $\chi_\bullet \ll 1$).
    
    In the bottom panel of Fig. \ref{fig:detRate}, we use the more realistic, non--uniform (in SMBH mass) TDE rate given by Eq.~\ref{eq:tdeRate}.  Smaller SMBH masses are strongly favored in flux--limited X--ray TDE samples, because (i) their disks have higher effective temperatures, increasing the luminosity in the eROSITA band; (ii) they preferentially occur in denser and cuspier galactic nuclei, where two--body relaxation times are shorter and TDE rates are higher; (iii) such SMBHs are more common, given our assumed mass function. Our predictions are closest to those of \citet{2014MNRAS.437..327K} when we set $\chi_\bullet\approx 0.9-0.95$, where the effective temperature at the inner disk edge in our model matches theirs.
    The observed black hole mass distribution of soft X--ray selected TDEs (\citealt{wevers-masses-ii}) does not show evidence for a larger number of TDEs from smaller SMBH masses, although there is a hint for this in hard X--ray selected TDEs.  
 
    Table~\ref{tab:rates} shows the mass--integrated eROSITA detection for a few different SMBH spin parameters (assuming 
    equal intrinsic numbers of prograde and retrograde disruptions). The detection rate is a strong function of the uncertain SMBH spin distribution: in our fiducial model (where the SMBH mass function extends down to $M_\bullet = 10^5 M_\odot$) we predict $\approx$1000 detections per year for $\chi_\bullet$=0.99, but only 170 per year for $\chi_\bullet=0$. For large values of $\chi_\bullet$, a flux-limited sample is strongly dominated by the $50\%$ of TDE disks we assume to align into prograde equatorial configurations; depending on the combination of $M_\bullet$ and $\chi_\bullet$, this ``prograde bias'' can range from $\sim 10\%$ to multiple orders of magnitude.  Prograde disks and high values of $|\chi_\bullet|$ are favored because of their higher bolometric luminosities and effective temperatures.
    
    The X--ray discovery rate is dominated by the smallest ($M_\bullet \sim 10^5 M_{\odot}$) SMBHs, a part of parameter space where the SMBH occupation fraction is poorly constrained. Interestingly, a flux--limited and X--ray selected TDE sample can be a more sensitive probe of the bottom end of the SMBH mass function than a volume--complete TDE sample would be\footnote{Using Eq.~\ref{eq:tdeRate}, we find that reducing $M_{\rm min}$ from $10^6 M_\odot$ to $10^5 M_\odot$ increases the volumetric TDE rate by a factor $\approx 8.5$.}.  In our models, this is true for $\chi_\bullet \lesssim 0.9$, and is due to the fact that (unless most SMBHs are nearly extremal) X--ray emission is typically on the Wien tail of TDE disks, and is thus highly sensitive to populations of smaller SMBHs.  Furthermore, the eROSITA TDE detection rate may also be a strong indicator of the SMBH spin distribution, even if the spins of individual TDE--hosting SMBHs cannot be measured.  This is analogous to the manner in which statistical samples of TDEs may probe the SMBH spin distribution near the Hills mass \citep{Kesden12}, although not limited to the largest TDE hosts.
    
    Many TDE light curves would be reasonably well--sampled with eROSITA. For example, for a $10^5$ ($10^6$) $M_{\odot}$ SMBH with a spin of 0.99, prograde disruptions would be visible on average for 27 (5.3) years. This would give an average of eight detections per TDE, considering the cadence and nominal duration of the eROSITA all--sky survey (six months and four years respectively). 

\begin{table*}[]
    \caption{Estimated eROSITA TDE detection rates.}
    \centering
    \begin{tabular}{c|ccc|ccc}
     & & $\dot{N} (M_{\bullet}\geq 10^5 M_{\odot})$
  &  & & $\dot{N} (M_{\bullet}\geq 10^6 M_{\odot})$  &   \\
  \hline
$\chi_{\bullet}$  & Total & Retrograde & Prograde & Total &Retrograde&Prograde\\
  & [yr$^{-1}$]& [yr$^{-1}$] & [yr$^{-1}$]& [yr$^{-1}$]&  [yr$^{-1}$] & [yr$^{-1}$] \\
    \hline
      0   & 172.3 & --& --&  4.8 &--& --\\
      0.1 & 174.3 & 76.0 & 98.3  & 5.0 & 3.1 &  1.9\\
      0.5 & 232.3 & 48.3 & 184.0  &  10.8 & 0.8 &  10.0\\
      0.9 & 551.3 & 32.6 & 518.7  & 65.8 & 0.4 &  65.4 \\
      0.99& 992.9 & 29.9 & 963.0  & 192.6 & 0.3 &  192.3 \\
    \end{tabular}
    \label{tab:rates}
    \tablecomments{
    Estimated eROSITA TDE detection rates using the formalism outlined in $\S$~\ref{sec:theoryRates}. The first column is the SMBH spin. Columns 2--4 give the total, retrograde, and prograde detection rates including all SMBHs between $10^5$  $M_{\odot}$ and the Hills mass. 
    Columns 5--7 give the total, retrograde, and prograde detection rate including SMBHs with masses between $10^6$ and the Hills mass. In all cases we assumed an equal intrinsic number of prograde and retrograde disruptions (see the discussion in $\S$~\ref{sec:retro}). For these estimates, we assume the TDE mass function from Eq.~\eqref{eq:tdeRate}, but discard TDEs with Galactic latitudes $\leq$30$^{\circ}$, as in \citet{2014MNRAS.437..327K}.}
\end{table*}

\subsubsection{Model II}
\label{sec:empiricalRates}
Next, we re--estimate eROSITA detection rates with a quasi--empirical model calibrated to reproduce the observed properties of \srctwo. While the model from the prior section was arguably an optimistic one (in its assumption that all TDEs will become X--ray bright after the disk accretion rate becomes sub-Eddington), this empirically calibrated model can be viewed as a rather pessimistic scenario, where we impose a long, adjustable period of X--ray darkness on most TDEs.  In this model, the bolometric luminosity is
\begin{align}
    &L_{\rm bol}=
    \begin{cases}
        0 &,\,t\leq t_{\rm o}\\
        \min\Bigl[L_{\rm Edd}(M_\bullet),2.5\times 10^{43} {\rm erg\,\,s^{-1}}\nonumber\\
         \left(\frac{t}{t_{\rm fall}}\right)^{-1.2} \left(\frac{M_\bullet}{3\times 10^6 M_{\odot}}\right)^{-1/2}\Bigr] &,\, t\geq t_{o}.
    \end{cases}\nonumber\\
    &t_o=\max[t_{\rm br}, t_{\rm Edd}, t_{\rm fall}]
    \label{eq:empirical_lum}
\end{align}
This reproduces the inferred late--time bolometric luminosity for \srctwo\footnote{$2.7\times 10^{41}$ erg s$^{-1}$ derived from the best fit black body spectrum for this event.} for its inferred SMBH mass of $\sim 3\times 10^6 M_{\odot}$ \citep{wevers-masses-ii}. The scalings with SMBH mass and time are the same as in the theoretical model, but SMBH spin is not explicitly included.  The X--rays turn on after the brightening time ($t_{\rm br}$), as long as this is greater than the fallback time and the luminosity is sub--Eddington. 

We assume, based on our late--time observations of \srctwo{}, that the spectrum is a blackbody with effective temperature $k T=$ 0.18 keV. The bolometric luminosity is up to two orders of magnitude smaller than in Model I, which would reduce the detection rate.  However, this is partially compensated for by the harder spectrum. 

Table~\ref{tab:rates2} shows the mass--integrated eROSITA detection rates for two different brightening times $t_{\rm br}$. For higher SMBH minimum masses ($M_{\rm min} = 10^6 M_{\odot}$) and small brightening times, the rates are comparable to the
estimates from Model I for moderate spins (with $0.5 \lesssim \chi_\bullet \lesssim 0.9$--see Table~\ref{tab:rates}). However, the predicted rates for lower SMBH mass limits ($M_{\rm min} = 10^5 M_{\odot}$) and/or larger brightening times are smaller than the zero spin case in Model I. 

\begin{table}[]
    \caption{Estimated TDE detection rates with 
    eROSITA using the quasi-empirical model outlined in $\S$~\ref{sec:empiricalRates}.}
    \centering
    \begin{tabular}{c|c|c}
    $t_{\rm br}$ & $\dot{N} (M_{\bullet}\geq 10^5 M_{\odot})$
  & $\dot{N} (M_{\bullet}\geq 10^6 M_{\odot})$
  \\
  yr & [yr$^{-1}$] &  [yr$^{-1}$] \\
    \hline
      1 &  150 & 24 \\
      5 & 16 & 2.6
    \end{tabular}
    \label{tab:rates2}
\end{table}

Taking into account that eROSITA will require the detection of at least 40 counts whereas on a previous pass eROSITA did not detect the source, in Model II, TDEs are on average observable for 2 (6) years after detection for a brightening time of 1 (5 years).  This implies significantly poorer temporal sampling in eROSITA light curves than in Model I.

\subsection{Retrograde and prograde TDE disks}
\label{sec:retro}
In the previous section, we saw that X--ray selected TDE samples are likely to possess a strong bias towards prograde configurations of SMBH spin $\vec{\chi}_\bullet$ and disk angular momentum $\vec{L}_{\rm d}$ (i.e. $\vec{\chi}_\bullet \cdot \vec{L}_{\rm d}>0$), so long as typical SMBH spin magnitudes are reasonably large ($\chi_\bullet \gtrsim 0.5$).  In this section, we discuss prospects for observing this prograde preference, and build simple toy models to show how it contrasts with the likely weaker orientation biases in optically selected TDE samples, which may even exhibit a preference for retrograde configurations ($\vec{\chi}_\bullet \cdot \vec{L}_{\rm d}<0$)

Our observations of \srctwo{} and observations presented in the literature for IC~3599 indicate that soft X--ray emission {\it may} remain visible for roughly a decade after the peak of a tidal disruption flare, although the majority of X--ray observations  of TDEs done $>8$ years after discovery find that they have transitioned to a non--thermal state with powerlaw spectral shapes where the power law index is $\approx2.5$. Nevertheless, for those sources with a soft spectrum years after the initial discovery, one could use such late--time TDE observations to directly measure SMBH spin through continuum fitting techniques.  While continuum fitting is a fruitful method of measuring the spins of stellar--mass black holes in XRBs (\citealt{Jeff2014SSRv}), it has only rarely been applied to SMBHs because{\it (i)} AGN typically produce dusty tori, and these complicate the X--ray spectral fitting, {\it (ii)} the spectral peak of a quasi--thermal AGN disk is usually in observationally inaccessible EUV bands.

The relatively cool temperatures of TDE disks (in contrast to those of XRBs) mean that quasi--thermal soft X--rays will generally be on the Wien tail of emission \citep{Lodato&Rossi11}, and their production will be dominated by the innermost gas annuli of the disk.  As a result, quasi--thermal X--rays from TDEs will be exponentially sensitive to the size of the disk inner edge, and therefore will depend strongly on SMBH spin.  At early times, the disk inner edge is nontrivial to estimate.  Because two--body scattering feeds stars to SMBHs from a roughly isotropic distribution of angles, TDE disks are generically born with order unity tilts.  A tilted thin disk will be truncated near the innermost stable spherical orbit (ISSO), but the high early--time accretion rates of TDEs may cause their innermost disk annuli to extend inside the ISSO\footnote{For example, with accretion disks tilted with respect to the black hole spin by an angle $15^\circ$ and a thickness of the order of 0.2, the simulations of \citet{Fragile2009} found the inner edge to be nearly independent of spin.}.  A greater problem at early times, however, is the messy hydrodynamical environment of the disk: if the stellar pericenter was sufficiently non--relativistic ($R_{\rm p} \gg R_{\rm g}$), the disk may retain substantial eccentricity \citep{Shiokawa+15}, and if optically thick stellar debris subtends a large solid angle on the sky, the majority of the X--ray flux may be absorbed in a reprocessing layer \citep{Guillochon+14, Metzger&Stone16}.

At late times, however, accretion rates will have dropped to sub-Eddington levels, shifting the disk inner edge close to the test particle value; many fallback times will have passed, enabling more complete circularization \citep{Hayasaki+16, Bonnerot+17}; reprocessing layers will have dissipated, revealing the inner disk \citep{Metzger&Stone16, vanVelzen+18}; and internal torques will have had time to align the TDE disk angular momentum vector with the black hole spin vector \citep{Franchini+16}.  Thus, if we interpret the soft X-ray spectrum of \srctwo{}, and perhaps \srcthree{} as quasi--thermal, it is reasonable to expect thin disks in the SMBH equatorial plane, with inner edges at the test particle ISCO.  We may now ask the question: do we expect an imbalance in the number of prograde and retrograde TDEs?  For a volume--complete sample the answer is clearly no.  However, for a more practical, flux--limited, sample of TDEs, there are strong reasons to suspect an imbalance.  We have already predicted that flux-limited, X-ray selected TDE samples can exhibit an enormous prograde bias (e.g. Table \ref{tab:rates}).  In this section, we investigate whether the same bias should be apparent for a flux-limited but optically selected TDE sample.  While the origin of TDE optical emission remains contested between ``shock-powered'' and ``reprocessing'' models, both of these scenarios have peak luminosities that will depend strongly on the orbital precession of debris streams, and therefore on SMBH spin.

To leading post-Newtonian (PN) order, both apsidal precession (precession of the debris stream's Runge--Lenz vector within the orbital plane) and nodal precession (precession of the orbital plane's angular momentum vector about the SMBH spin vector) are larger for retrograde than for prograde orbits (\citealt{Merritt2010}).  Neglecting for now the possibility that extreme nodal precession may prevent stream self-intersections\footnote{Tidal disruptions of stars in the relativistic regime (e.g.~white dwarfs disrupted by intermediate--mass BHs, or solar mass main sequence stars disrupted by a BH with $M_\bullet=10^{7-8}$\msun) around spinning SMBHs, may lead to stellar debris streams that fail to promptly self--interact, unless the inclination of the stellar orbit is nearly perpendicular to the BH spin axis or if the thickness of the debris streams is large enough such that they always intersect (\citealt{dai2013}; \citealt{Guillochon&RamirezRuiz15}, \citealt{Hayasaki+16}).}, the greater apsidal shifts for debris from retrograde TDEs means that these debris streams will self-intersect and dissipate energy at smaller radii. Smaller stream self-intersection radii $R_{\rm SI}$ will probably yield higher peak optical luminosities, regardless of the dominant optical power source in observed TDEs.  In the ``reprocessing paradigm,'' smaller $R_{\rm SI}$ will mean faster disk formation and higher peak accretion rates $\dot{M}$ onto the SMBH, although this must be weighed against the potentially greater radiative efficiency of prograde disks.  In the ``circularization paradigm,'' smaller $R_{\rm SI}$ values will thermalize greater amounts of bulk kinetic energy.

The translation between self--intersection radius $R_{\rm SI}$ and optical luminosity is currently an unsolved problem.  Under the assumption that most of the observed optical emission is shock-powered, we will use the following toy model for peak luminosity:
\begin{equation}
L_{\rm peak} = \eta_{\rm SI} \frac{GM_\bullet \dot{M}_{\rm peak}}{R_{\rm SI}}, \label{eq:LPeakCirc}
\end{equation}
where, as before, $\dot{M}_{\rm peak}$ is the peak mass fallback rate.  The dimensionless number $\eta_{\rm SI}\le 1$ is the fraction of stream kinetic energy thermalized {\it and} radiated at the self-intersection; for simplicity, we take it to be a constant\footnote{As \citet{Lu&Bonnerot19} have noted, a large fraction of the thermalized stream kinetic energy may be lost to adiabatic degradation prior to the time it can be emitted.  Because the fractional energy loss to $P{\rm d}V$ work depends on gas optical depth at $R_{\rm SI}$ and therefore on $M_\bullet$ and other parameters, the assumption of constant $\eta_{\rm SI}$ is crude.  Deriving a more complete theoretical model is, however, beyond the scope of this work.}.  A flux-limited survey will find a differential number of TDEs per bin of pericenter $R_{\rm p}$ and inclination $i$ that scales as ${\rm d}N_{\rm det}/{\rm d}i{\rm d}R_{\rm p} \propto L_{\rm peak}^{3/2}(i, R_{\rm p})({\rm d}\dot{n}/{\rm d}i{\rm d}R_{\rm p})$, where the differential rate ${\rm d}\dot{n}/{\rm d}i{\rm d}R_{\rm p} \propto \sin i$ if we are in the full loss cone (FLC) regime, and ${\rm d}\dot{n}/{\rm d}i{\rm d}R_{\rm p} \propto \sin i\times \delta(R_{\rm p}-R_{\rm t})$ if we are in the empty loss cone (ELC) regime (we have assumed isotropy in stellar arrival directions in both regimes).

The dependence of $L_{\rm peak}$ on $i$ and $R_{\rm p}$ can be computed by defining the self-intersection radius \citep{Dai+15}
\begin{equation}
    R_{\rm SI} = \frac{R_{\rm p}(1+e)}{1+e\cos(\pi + \delta \omega / 2)},
\end{equation}
where $e$ is the eccentricity of the elliptical orbit of the stream of material formed by the disrupted star.  For convenience, we take the eccentricity of the most tightly bound debris, $e_{\rm min} = 1-2(M_\star / M_\bullet)^{1/3}/\beta$.
Here, we have made use of the per--orbit apsidal shift angle, $\delta\omega$, which, to lowest PN order in dimensionless SMBH spin, $\chi_\bullet$, is \citep{Merritt2010} 
\begin{equation}
    \delta\omega = A_{\rm S}-2A_{\rm J}\cos i ,
\end{equation}
where
\begin{align}
    A_{\rm S} =& \frac{6\pi}{c^2} \left( \frac{GM_\bullet}{R_{\rm p}(1+e)} \right) \\
    A_{\rm J} =& \frac{4\pi \chi_\bullet}{c^3} \left( \frac{GM_\bullet}{R_{\rm p}(1+e)} \right)^{3/2}.
\end{align}
In the empty loss cone regime, for fixed SMBH mass and stellar properties, the retrograde fraction is simply
\begin{equation}
    f_{\rm ret}^{\rm ELC} = \frac{\int^\pi_{\pi/2}\sin i[1+e\cos(\pi+\delta\omega/2)]^{3/2}{\rm d}i}{\int_0^\pi\sin i[1+e\cos(\pi+\delta\omega/2)]^{3/2}{\rm d}i},
\end{equation}
  
In the full loss cone (FLC) regime, a second integral is necessary:
\begin{equation}
    f_{\rm ret}^{\rm FLC} = \frac{\int^\pi_{\pi/2} \sin i \int^{R_{\rm t}}_{R_{\rm min}}R_{\rm p}^{-3/2}[1+e\cos(\pi+\delta\omega/2)]^{3/2}{\rm d}R_{\rm p}{\rm d}i}{\int_0^\pi\sin i \int^{R_{\rm t}}_{R_{\rm min}} R_{\rm p}^{-3/2} [1+e\cos(\pi+\delta\omega/2)]^{3/2}{\rm d}R_{\rm p}{\rm d}i}.
\end{equation}
Here $R_{\rm p}$ ranges from a maximum value of $R_{\rm t}$ down to a minimum value of $R_{\rm min}(\chi_\bullet, i)$.  This minimum value, the innermost bound spherical orbit (IBSO), is computed from the Kerr geodesic equations \citep{Bardeen+72}.  The IBSO is larger for retrograde spins, which (via Eq.~\ref{eq:LPeakCirc}) introduces a prograde bias.

\begin{figure}
\includegraphics[width=85mm]{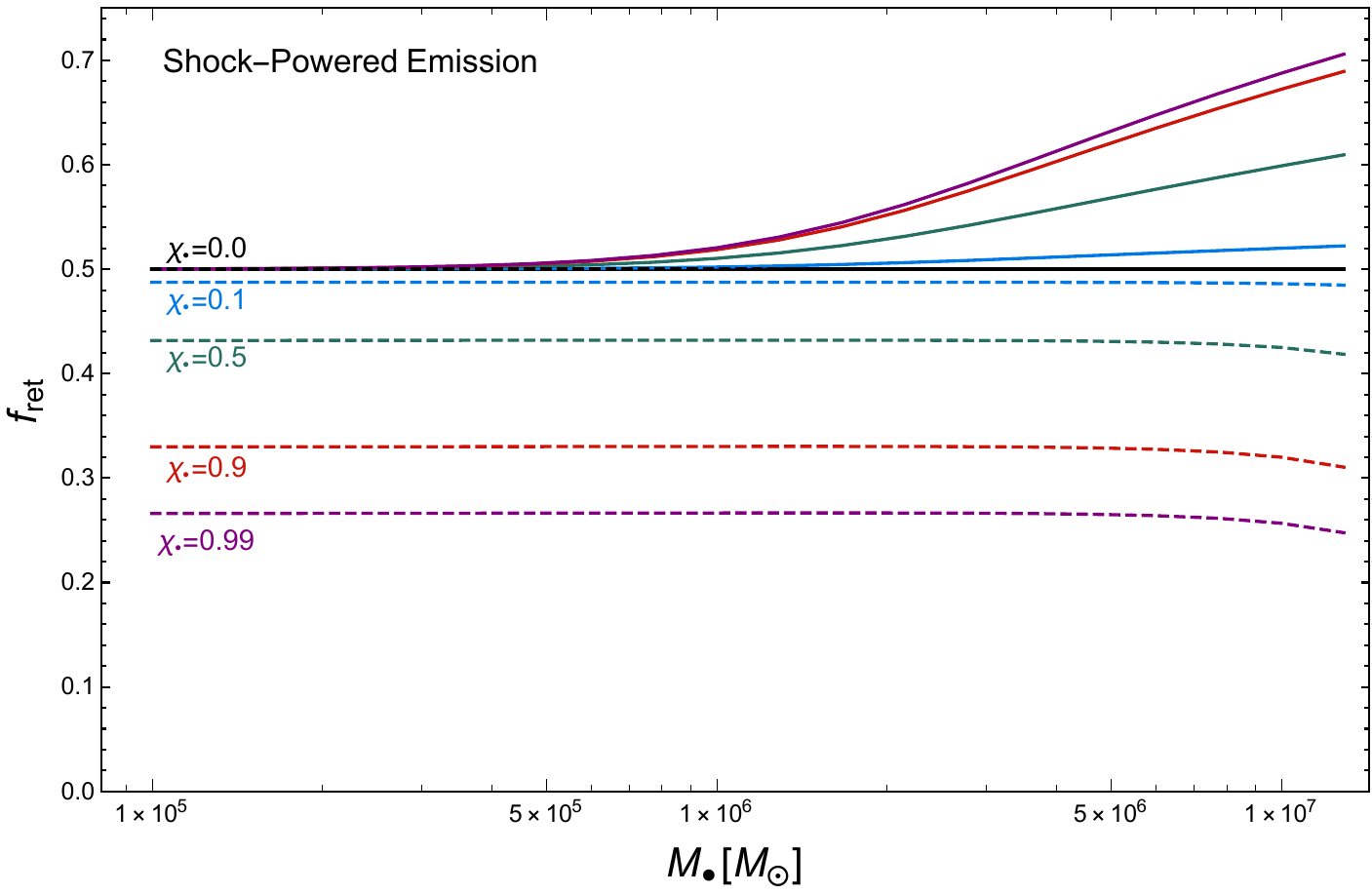}
\caption{The fraction of TDEs with retrograde disks, $f_{\rm ret}$, in a flux-limited sample of (i) optically--selected and (ii) shock-powered tidal disruption flares.  In the empty loss cone regime (solid lines), there is no preference for retrograde orbits when SMBH spin $\chi_\bullet$ is zero (the Schwarzschild limit), but the preference becomes more notable for higher values of $\chi_\bullet$ (shown and labeled as color-coded curves).  Conversely, in the full loss cone regime (dashed lines), the preference is for prograde disks.}
\label{fig:retroBiasCirc}
\end{figure}

We illustrate the overall orientation bias in flux--limited samples of shock--powered TDEs in Fig.~\ref{fig:retroBiasCirc}.  There is no bias when $\chi_\bullet=0$ (as symmetry demands), but the bias becomes notable when $\chi_\bullet \gtrsim 0.5$.  Interestingly, the bias is qualitatively different in the two regimes of loss cone repopulation.  In the empty loss cone regime, there is almost no bias for $M_\bullet \lesssim 10^6 M_\odot$, but a moderate retrograde bias for larger SMBHs.  In the full loss cone regime, there is a moderate prograde bias across all bins of $M_\bullet$.  Since the empty loss cone regime predominates for high--mass SMBHs, and the full loss cone regime predominates for low--mass SMBHs \citep{stone&metzger2016}, we expect that flux--limited, shock--powered TDE samples will exhibit a prograde bias for $M_\bullet \lesssim 10^{6.5} M_\odot$, and a retrograde bias at higher masses.

\begin{figure}
\includegraphics[width=85mm]{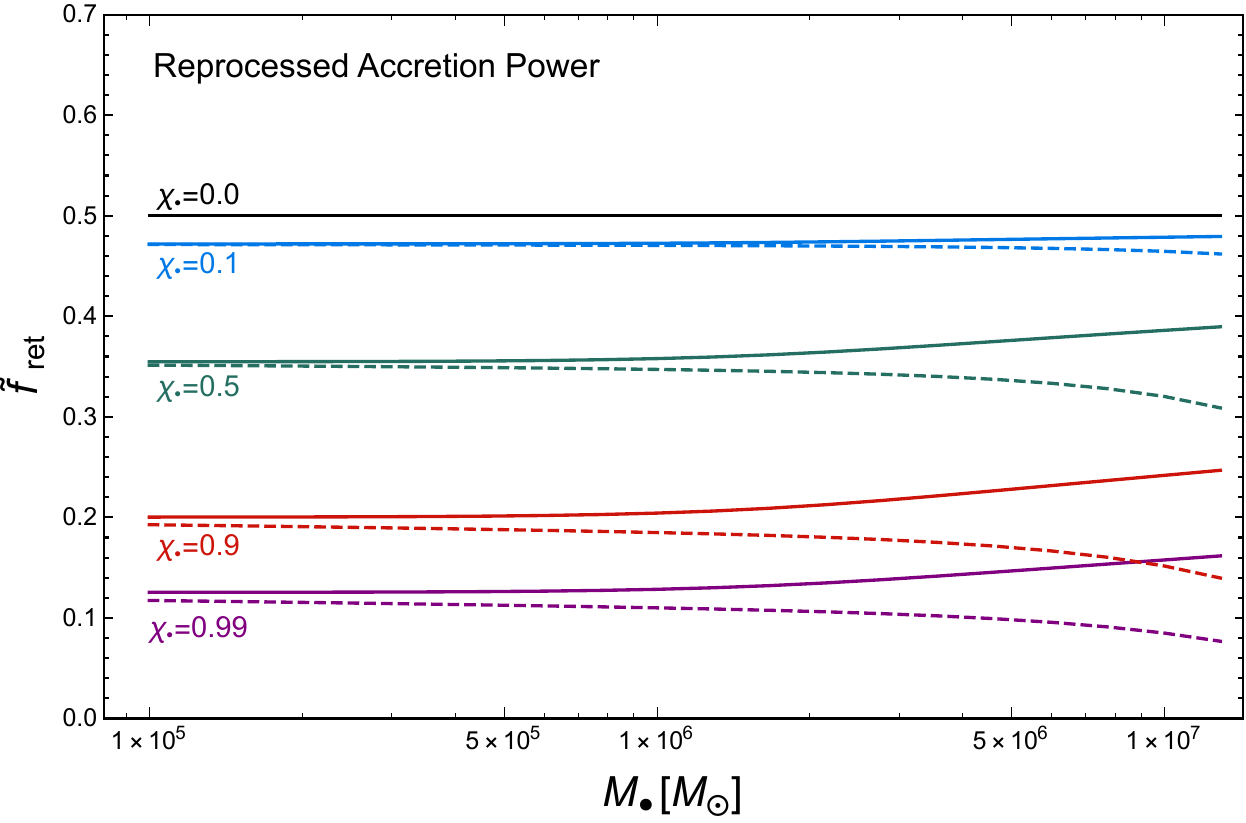}
\caption{Same as Fig. \ref{fig:retroBiasCirc}, but we now compute the retrograde fraction ($\tilde{f}_{\rm ret}$) considering a model for optical emission based on reprocessed X-ray/EUV emission from an inner accretion disk.  In contrast to the shock--powered model of Fig. \ref{fig:retroBiasCirc}, reprocessing-powered TDEs almost always show a bias for {\it prograde} disks due to radiative efficiency considerations.  This bias is generally strongest for the empty loss cone regime and smaller SMBHs, but depends somewhat on the power law index $\xi$ (assumed to be 0.5 in this plot).  If $\xi \gtrsim 2$, a weak retrograde bias may be recovered in the empty loss cone regime.}
\label{fig:retroBiasRep}
\end{figure}

We may also consider a similar sort of toy model for the reprocessing picture of TDE optical luminosity, designed to illustrate the competition between disk formation (which is faster for retrograde orbits) and the radiative efficiency of a circularized accretion disk (which is higher for prograde orbits).  Let us say that the peak optical luminosity in a reprocessing model is 
\begin{equation}
    L_{\rm peak} = \eta_\bullet \eta_{\rm r} \dot{M}_{\rm max}c^2,
\end{equation}
where $\eta_\bullet$ is the standard radiative efficiency of a thin, equatorial accretion disk (see $\S$~\ref{sec:rates}), and the efficiency with which an optically thick reprocessing layer converts X--ray and extreme UV photons to optical emission is assumed (again, for simplicity) to be a constant, $\eta_{\rm r}$.  Here $\dot{M}_{\rm max}$ does not represent the peak mass fallback rate to pericenter, $\dot{M}_{\rm max} =\frac{M_\star}{3}(t/t_{\rm fall})^{-5/3}$, but rather the peak accretion rate through the disk, which we parametrize as
\begin{equation}
    \dot{M}_{\rm max} = \frac{M_\star}{2t_{\rm circ}},
\end{equation}
where we assume that the ``circularization timescale'', $t_{\rm circ}$, is a function only of the self--intersection radius, and is related to the fallback time for the most tightly bound debris as $t_{\rm circ} = t_{\rm fall}(R_{\rm SI}/R_{\rm g})^\xi$.  This power--law parametrization of the disk formation timescale is crude, but will suffice to explore what types of disk orientation biases we expect if reprocessing is responsible for the observed optical emission.  We find modified versions of the empty and full loss cone regime retrograde fractions:
\begin{equation}
    \tilde{f}_{\rm ret}^{\rm ELC} = \frac{\int^\pi_{\pi/2}\sin i[1+e\cos(\pi+\frac{\delta\omega}{2})]^{3\xi/2}\eta_\bullet^{3/2}{\rm d}i}{\int_0^\pi\sin i[1+e\cos(\pi+\delta\omega/2)]^{3\xi/2}\eta_\bullet^{3/2}{\rm d}i},
\end{equation}
and
\begin{equation}
    \tilde{f}_{\rm ret}^{\rm FLC} = \frac{\int^\pi_{\pi/2} \sin i \int^{R_{\rm t}}_{R_{\rm min}}\left(\frac{\eta_\bullet}{R_{\rm p}^\xi}\right)^{3/2}[1+e\cos(\pi+\frac{\delta\omega}{2})]^{3\xi/2}{\rm d}R_{\rm p}{\rm d}i}{\int_0^\pi\sin i \int^{R_{\rm t}}_{R_{\rm min}} \left(\frac{\eta_\bullet}{R_{\rm p}^\xi}\right)^{3/2} [1+e\cos(\pi+\frac{\delta\omega}{2})]^{3\xi/2}{\rm d}R_{\rm p}{\rm d}i}.
\end{equation}
We illustrate the retrograde fractions in a flux--limited, reprocessing-powered TDE sample in Fig. \ref{fig:retroBiasRep}.  In contrast to our earlier shock--powered calculations, our toy model for reprocessing power almost always exhibits a {\it prograde} disk bias, as this configuration yields much higher radiative efficiencies.  The detailed nature of the orientation bias depends on the power law index $\xi$ encoding the dependence of circularization efficiency on $R_{\rm SI}$ (in this figure, we use $\chi=0.5$).  Very high values of $\xi$ ($\gtrsim 2$) can create a retrograde bias in a sample of TDEs in the empty loss cone regime, but this level of sensitivity to $R_{\rm SI}$ is not suggested by existing hydrodynamical simulations of circularization \citep{Hayasaki+16, Bonnerot+16}.  The overall level of bias depends on $\chi_\bullet$, but shows little variation with $M_\bullet$.

Of the TDEs we have observed, both \srctwo{} and \srcthree{} have late--time accretion disks whose FUV properties were modeled in \citet{vanVelzen+18}.  Our X--ray detections are compatible with these disk models provided the disk in \srctwo{} is prograde with respect to a very rapidly spinning SMBH, and the disk in \srcthree{} is retrograde with respect to a spinning SMBH.  While this sample is too small (and our disk models so far too crude) to meaningfully constrain $f_{\rm ret}$, these observations, and the arguments in this section, highlight the potential of future late--time observations and modeling to determine the dependence of peak flare luminosity on the inclination of the disrupted star's orbit.  This may also serve as a useful test between different models of optical power sources in TDEs, as it would be hard to explain a pronounced retrograde bias in the reprocessing paradigm.

In this section, we have used simple but illustrative optical emission models to demonstrate that the selection effects operating in flux-limited, {\it optically selected} TDE samples favor a very different $\chi_\bullet$ distribution than do the selection effects in flux-limited, X--ray selected samples (\S \ref{sec:rates}).  Specifically, an X--ray selected sample will be biased strongly towards prograde orbits around rapidly spinning SMBHs, while an optically selected sample will still be biased towards high $|\chi_\bullet|$, but much more weakly so, and may possess either a  prograde or retrograde bias depending on the specific optical emission mechanism.  A consequence of this is that the quasi-thermal X--ray luminosities in optically selected TDE distributions should be systematically lower than the corresponding X--ray luminosities in an X--ray selected sample, since the former will have cooler disk temperatures, on average.  

\section{Conclusions}
We have conducted \chan ~X--ray observations of four optically--selected TDEs long after the peak of their optical flares.  In three cases we detected late--time soft X--ray emission: \srcone{}, \srctwo{}, and \srcthree{} are best--fit with unabsorbed ($0.3-7~{\rm keV}$) luminosities of $(3.2\pm0.2)\times 10^{42}~{\rm erg~s}^{-1}$, $3.9^{+1.1}_{-1.0}\times 10^{41}~{\rm erg~s}^{-1}$, and $9^{+9}_{-5}\times 10^{40}~{\rm erg~s}^{-1}$, respectively.  Our fourth target, \srcfour{}, was undetected by \chan{}, yielding an upper limit on its soft X--ray luminosity of $L_{\rm X} < 3 \times 10^{41}~{\rm erg~s}^{-1}$.  Three of these observations represent the longest temporal baseline for X--ray observations of optically--selected TDEs to date: \srcone{} and \srctwo{} were observed roughly eight years after peak, while \srcfour{} was observed roughly ten years post--peak.

These TDEs exhibit a diversity of X--ray behavior at late times.  The X--ray spectrum of \srctwo{} is best fit by a very soft power law (and has similarities to the high--soft state of XRBs).  In contrast, the X--ray spectrum of \srcone{} is best fit as a comparatively hard, non--thermal power law, quite unlike most TDEs seen at early times, but not unlike several other TDEs seen at similarly late times as shown in the literature, and more similar to the spectrum of an AGN or the low--hard state of an XRB.  \srcthree{} does not have sufficient X--ray counts to determine the shape of its spectrum.

Our primary conclusions are as follows:
\begin{enumerate}
    \item Late--time X--ray detections are further evidence that \srcone{}, \srctwo{}, and, to a lesser extent, \srcthree{} represent bonafide TDEs and not a peculiar type of nuclear supernova explosion.  The persistence of high X--ray luminosities $\approx 5-10$ yr post-peak also argues strongly against the presence of a thermal instability in TDE disks, as would be predicted by simple $\alpha$-disk theory.  
    \item We hypothesize that the marked spectral differences between \srcone{} and \srctwo{} may have been caused by a late--time state change in \srcone{}\,to a low--hard state (in analogy to the state changes regularly observed in black hole X--ray binaries). Radio follow--up observations of \srcone\, could test this hypothesis, as could continued X--ray monitoring of \srctwo{} to investigate if it also exhibits a state change to a (harder) power-law spectrum.
    \item Assuming that our observations 4--9 years after optical detection are not caused by short--lived flares, we conclude that most TDEs are persistently bright X--ray sources visible for at least a decade. This is in line with previous work. Overall, nine out of ten TDEs observed in X--ray at late times ($\approxgt 5$ yr after its discovery) were detected. This has implications for detection
    rates in near future, wide field X--ray surveys if deep enough early-time X--ray observations of optically--selected TDEs exist. For example, we find the eROSITA instrument recently launched on the {\it Spectrum R\"ontgen Gamma} satellite could detect up to $1000$ TDE flares per year if most low mass SMBHs have near maximal spins. However, the detection rate would be 170 per year if most SMBHs have zero spin, and (in the Schwarzschild limit) would be further reduced to only 5 per year if SMBHs with masses below $10^6 M_{\odot}$ are excluded. 
    \item We propose that there is often a delay between the peak optical and the X--ray emission in TDEs, such that optical and X--ray selected TDEs are, in many cases, the same type of flare observed at different stages. For example, in X--ray selected TDEs the optical emission (e.g.~from the circularisation shock), may have already subsided  below the level that can be detected above the nuclear region of the host galaxy. 
    \item The persistence of a soft X--ray spectrum at late times (such as in \srctwo{}) opens up the possibility of black hole spin determinations using continuum fitting techniques (Wen et al.~in prep.). These were, in the past, primarily applied to black holes in soft--state X--ray binaries (\citealt{Jeff2014SSRv}).  Late--time X--ray observations will avoid, or at least minimize, theoretical uncertainties associated with early--time TDE disk modeling, such as generic disk tilts, significantly non--circular gas flows, and the presence of optically thick stellar debris on larger scales.  The number of X--ray photons detected in the current observations of \srctwo{} are, however, insufficient to attempt this exercise.
    \item If the SMBHs responsible for TDEs possess appreciable spins, a flux--limited sample of TDEs will generally be biased towards an excess of prograde or retrograd disks.  In an optically selected sample, the sign of this bias depends on the exact emission mechanism.  Shock-powered optical emission \citep{Piran+15} will exhibit a mild retrograde bias in the empty loss cone regime, and a mild prograde bias in the full loss cone regime. If instead, the optical emission is powered by reprocessed X--rays generated from a veiled inner accretion flow \citep{Guillochon+14, Metzger&Stone16}, then prograde black hole spins are almost always favored, usually by a factor of a few. X--ray selected TDE samples have a very strong (one--to--two orders of magnitude) bias for prograde orbits if most SMBHs are spinning rapidly. 
\end{enumerate}

\section*{Acknowledgments} We would like to thank the {\it Chandra X-ray Observatory} for approving and carrying out the observations presented in this paper and the anonymous referee for her/his comments which helped improve the paper.  \noindent PGJ acknowledges funding from the European Research Council under ERC Consolidator Grant agreement no 647208 and discussions with Giacomo Cannizzaro. NCS and AG acknowledge funding from {\it Chandra} GO 18700591.  NCS acknowledges additional funding from {\it Chandra} GO 20700515.  BDM acknowledges support from NASA through the Astrophysics Theory Program (grant number NNX17AK43G). This research has made use of the NASA/IPAC Extragalactic Database (NED), which is operated by the Jet Propulsion Laboratory, California Institute of Technology, under contract with the National Aeronautics and Space Administration.

\section*{Software} Tool 1 XSPEC version 12.10.1 (HEASOFT; \citealt{2014ascl.soft08004N}). Tool 2 Ciao version 4.10 developed by the \chan\, X--ray Center \citealt{2006SPIE.6270E..1VF}.

\bibliographystyle{aasjournal} 
\bibliography{papers}

\end{document}